\documentclass[a4paper,fleqn,usenatbib]{mnras}

\usepackage{graphicx}	
\usepackage{amsmath}	
\usepackage{amssymb}	
\usepackage{multicol}        
\usepackage{bm}		
\usepackage{pdflscape}	

\usepackage[T1]{fontenc}
\usepackage{ae,aecompl}

\usepackage{newtxtext,newtxmath}

\title[500 micron Risers III]{The Nature of 500 micron Risers III: A Small Complete Sample}
\author[D.L. Clements et al.]{
D.L. Clements$^{1}$\thanks{Contact e-mail: d.clements@imperial.ac.uk},
J. Cairns$^{1}$,
J. Greenslade$^{1}$,
G. Petitpas$^{2}$, Y. Ding$^{1}$,\newauthor
I.~P{\'e}rez-Fournon$^{5,6}$
D. Riechers$^{7,8,9}$
\\
$^{1}$Imperial College London, Prince Consort Road, London SW7 2AZ, UK\\
$^{2}$Harvard-Smithsonian Center for Astrophysics, 60 Garden Street, Cambridge, MA 02138 \\ 
$^3$Department of Physics and Astronomy, University of British Columbia, 6224 Agricultural Road, Vancouver, BC V6T-1Z1, Canada\\
$^{4}$Center for Astrophysics and Space Astronomy 389-UCB, University of Colorado, Boulder, CO, 80309, USA\\
$^{5}$Instituto de Astrof\'{i}sica de Canarias, C/ V\'{i}a L\'{a}ctea, E-38200 La Laguna, Tenerife, Spain\\
$^{6}$Departimento de Astrof\'{i}sica, Universidad de La Laguna, E-38206, La Laguna, Tenerife, Spain\\
$^7$	Cornell University, 220 Space Sciences Building, Ithaca, NY 14853, USA\\
$^8$Max-Planck-Institut f\"ur Astronomie, K\"onigstuhl 17, D-69117 Heidelberg, Germany\\
$^9$Humboldt Fellow
}\date{}
\pagerange{\pageref{firstpage}--\pageref{lastpage}}
\pubyear{}

\begin{document}

\maketitle

\label{firstpage}

\begin{abstract}
{\em Herschel} surveys have found large numbers of sources with red far-IR colours, and spectral energy distributions (SEDs) rising from 250 to 500$\mu$m: 500 risers. The nature and role of these sources is not fully understood.  We here present Submillimeter Array (SMA) interferometric imaging at 200 GHz of a complete sample of five 500 risers with F500 $>$ 44 mJy selected within a 4.5 square degree region of the XMMLSS field. These observations can resolve the separate components of multiple sources and allow cross identification at other wavelengths using the extensive optical-to-IR data in this field. Of our five targets we find that two are likely gravitationally lensed, two are multiple sources, and one an isolated single source. Photometric redshifts, using optical-to-IR data and far-IR/submm data, suggest they lie at redshifts $z \sim 2.5 - 3.5$. Star formation rates and stellar masses estimated from the SEDs show that the majority of our sources lie on the star-formation rate-stellar mass `main sequence', though with outliers both above and below this relation. Of particular interest is our most multiple source which consists of three submm emitters and one submm-undetected optical companion within a 7 arcsecond region, all with photometric redshifts $\sim$ 3. One of the submm emitters in this group lies above the `main sequence', while the optical companion lies well below the relation, and has an estimated stellar mass of 3.3$\pm 1.3 \times$10$^{11}$ M$_{\odot}$. We suggest this object is a forming brightest cluster galaxy (BCG) in the process of accreting actively star forming companions.

\end{abstract}


\section{Introduction}

The {\em Herschel Space Observatory} has revolutionised our knowledge of the far-IR sky in general, and of the far-IR/submm properties of galaxies in particular. Surveys using the Herschel-SPIRE instrument \citep{g10}, operating simultaneously at wavelengths of 250, 350 and 500$\mu$m, such as the Herschel Multitier Extragalactic Survey (HerMES, \citet{o12}) and the Herschel Astrophysical Terahertz Large Area Survey (H-ATLAS, \citet{e10}) have expanded our samples of far-IR/submm galaxies from hundreds to hundreds of thousands. An unexpected aspect of this work has been the discovery of a large population of high redshift dusty star-forming galaxies (DSFGs), lying at redshifts $\geq 2-3$. These sources are identified through their colours in the three bands of the SPIRE instrument. Dust emission in a typical galaxy spectral energy distribution (SED) usually peaks at wavelengths around 100$\mu$m. As such a source moves to higher redshifts this peak is observed to move through the SPIRE bands in such a way that a source at a redshift of $\sim 3 - 4$ will have an observed SED that peaks in the 500$\mu$m band. Candidate high redshift sources can thus be identified as red objects in the SPIRE surveys, with SEDs peaking at 500 $\mu$m. These sources are generally known as 500 risers since their SEDs rise in the SPIRE bands from 250 to 500$\mu$m. Examples of such sources spectroscopically confirmed to lie at high redshift include HFLS3 \citep{r13}, at $z=6.34$ and ADFS-27 at $z=5.655$ \citep{r21}. Such far-IR/submm colour selection methods may also be extended to higher redshifts with the addition of longer wavelength observations, looking for so-called SPIRE dropout sources (eg. \citet{g19,y22}).

A small number of bright, often gravitationally lensed, 500 risers have been studied in detail, but the nature of the underlying population remains unclear. However, such sources are known to be quite numerous. Searches for 500 risers in the HerMES \citep{d14} and HELMS (HerMES Large Mode Survey, \citep{a16} surveys have found unexpectedly large numbers of such sources, amounting to an areal density of a few per square degree down to 500$\mu$m fluxes of 30 to 50mJy. To be detectable at these fluxes at such high redshifts these sources would have to be highly luminous in the rest-frame far-IR and thus forming stars at a very high rate $\geq$ 1000 M$_{\odot}$/yr. Such a population essentially does not exist in current models of galaxy formation and evolution (\citet{b05, l10, gr11, h11, h15, l16}) with the observed numbers tens to hundreds of times greater than most predictions. It has been suggested that such extreme DSFGs contribute significantly to the star-formation rate density of the universe (eg. \citet{rr16}), and that they are the high redshift progenitors of the massive red sequence galaxies that dominate local galaxy clusters (eg. \citet{o17}).

Explanations for the large number of 500 risers seen in {\em Herschel} surveys include the possibility that a substantial fraction of this population is strongly lensed \citep{n10} and the suggestion that many of the SPIRE sources are in fact multiple objects blended together by the large {\em Herschel} beams (\citet{b17, c17, d18}) with multiplicity rates as high as 98\% suggested. Followup observations of 500 risers have been undertaken at a variety of wavelengths to test these suggestions and others. Spitzer observations using IRAC at 3.6 and 4.5 $\mu$m \citep{m19} found that 65\% of the sources observed were unlensed and 27\% were multiple. Observations at 1.1mm with the AzTEC camera on the Large Millimetre Wave Telescope (LMT, \citet{m21}) found that 9\% of sources are multiple when imaged on scales of 9 arcseconds, with some having photometric properties suggesting that they are physically associated. Interferometric observations using the SMA \citep{g20} on scales of 0.35 to 3 arcseconds found that sources were likely to be multiple in 35 to 40\% of cases and that lensing played less of a role than predicted by some authors (eg. \citet{b17}) who have suggested lensed sources should be the dominant population at 500 $\mu$m fluxes $>$ 60mJy. ALMA followup at angular resolutions $\sim$0.12 arcseconds of a sample of 44 500 risers \citep{o17} found a multiplicity rate of 39\%, including the spectacular $z=4.002$ red cluster core nicknamed the Deep Red Core \citep{o18} and that 41\% of their sources were subject to gravitational lensing to a greater or lesser extent.

These results indicate that there is a clear role for both lensing and multiplicity among the 500 riser population, but that the fraction of lensed and multiple sources are not as high as many of the models have suggested and that some fraction of non-lensed, non-multiple extreme galaxies remain. However, two things are lacking from many of these various followup programmes. Firstly, they are not statistically complete, but result from incomplete samples of objects not necessarily selected in a uniform statistical manner. Secondly, they largely lack data at other wavelengths from cross identifications with other datasets. This makes the properties of the underlying galaxies, including redshifts, star-formation rates and stellar masses, difficult, if not impossible, to determine. The current study aims to address these problems by conducting $\sim$ 1 arcsecond resolution interferometric imaging of a small but statistically complete sample of 500 risers selected from the SERVS-XMM survey region \citep{n17} where an abundance of ancillary data at optical to mid-IR wavelengths exists, in addition to the Herschel data, and is readily available, together with detailed SED modelling using CIGALE \citep{z22}. With this small but complete sample we can start to draw statistically robust conclusions about the 500 riser population and their multiwavelength counterparts.

The rest of this paper is structure as follows. In the following section we discuss source selection and the available multiferquency data. In Section 3 we detail the SMA observations and their results. In Secion 4 we discuss the optical-to-mid-IR identifications of these SMA sources. In Section 5 we discuss the redshifts of these objects as obtained from the optical-to-near-IR SED fitting and from the far-IR data from {\em Herschel} and draw some conclusions regarding the lensing nature of some sources. We describe the nature of our sources and provide detailed notes on them in Section 6 before discussing these results in Section 7. We draw our conclusions in Section 8. Throughout this paper, we assume a concordance $\Lambda$CDM cosmology, with $H_0 = 67.74 $km s$^{-1}$ Mpc$^{-1}$, $\Omega_{\Lambda} = 0.69$, and $\Omega_m$ = 0.31.

\section{The SERVS-XMM 500 riser Sample}

The aim of this project is to use submillimetre interferometry observations from the SMA to pinpoint the locations of a sample of 500 risers chosen to lie in a region of the sky where plentiful existing multiwavelength data can be found. This will allow the optical to submm SEDs of these objects to be investigated. The HerMES survey \citep{o12} was devised in such a way that most of its survey fields are located in parts of the sky where there is plentiful multiwavelength data. Among these fields, the XMMLSS field, centred at 2:20:00 -04:40:00 and covering a total of 18.9 sq. degrees, is very well placed since it coincides with a number of other multiwavelength surveys include the SERVS deep 3.6 and 4.5 $\mu$m observations \citep{m12} with  the IRAC instrument on {\em Spitzer}. At the time this study was initiated a compendium of this photometry, produced using the psf matching code TRACTOR, was available over a 4.5 sq. degree region of the HerMES XMMLSS field \citep{n17}. Subsequent to this study beginning further compendiums of photometric data in this and other deep survey fields, together with SED modelling, has been undertaken by \citet{z22} which we later use as the basis for some of our analysis. In this section we first describe the selection of the 500 risers in this field from the HerMES catalogs, and then summarise the multiwavelength data available.

\begin{table*}
\begin{tabular}{ccccccc}\hline
HerMES ID&Short Name &RA&DEC&F250 (mJy)&F350 (mJy)& F500 (mJy)\\ \hline
  J021856.3-043541 & J021856&02:18:56.32 & -04:35:41.2 & 27.0$\pm$5  & 44.0$\pm$7 & 46.0$\pm$8 \\
  J022425.8-042740 & J022425&02:24:25.81 & -04:27:40.4 & 32.9$\pm$5  & 52.4$\pm$8 & 52.5$\pm$7 \\
   J022427.7-044923 & J022427&02:24:27.73 & -04:49:23.0 & 31.8$\pm$5  & 45.5$\pm$8 & 47.2$\pm$6 \\
  J022448.6-041223 &J022448& 02:24:48.56 & -04:12:23.6 & 46.0$\pm$5  & 57.4$\pm$8 & 59.2$\pm$6 \\
    J022610.5-050221 &J022610& 02:26:10.51 & -05:02:21.6 & 40.1$\pm$5 & 44.0$\pm$8  & 48.0$\pm$9 \\ \hline
\end{tabular}
\caption{Herschel positions and fluxes for the sources selected as 500 risers in the SERVS-XMM region.}
\label{table:sources}
\end{table*}

\subsection{Herschel Selection}

The XMMLSS field was observed by {\em Herschel} as part of the HerMES survey \citep{o12} at 250, 350 and 500$\mu$m, with 5$\sigma$ detection thresholds of $\sim$25.8, 21.2, and 30.8 mJy respectively. We use the standard (\citet{d14,a16}) method of selecting candidate high redshift luminous starbursts by looking for sources whose far-IR SED rises from 250 to 500$\mu$m. Existing followup of 500 risers (eg. \citet{r13, a16}) finds that all those with spectroscopic redshifts have z > 3 and most z > 4, in contrast to the wide variety of redshifts found for sources selected solely through 850$\mu$m flux (see eg. \citet{c05}). We find five 500 risers in the 4.5 sq. degrees of the SERVS-XMM field \citep{n17} down to a flux of 44mJy at 500$\mu$m, with fluxes ranging from 44 to 59 mJy (Table \ref{table:sources}). We choose the 44mJy flux limit at 500$\mu$m for several reasons: firstly, so as to have clear detections in the two other SPIRE bands despite these sources being 500 risers; and, secondly, to ensure that the sources will be bright enough for good detections by the SMA even if they are multiple sources, and finally to limit the sample size to a manageable number for SMA followup in a reasonable number of nights, given that each source will be observed for one track. The areal density of sources selected in this way is slightly over 1 per sq. degree, consistent with the areal density seen in the Dowell and Asboth 500 riser samples. It should be noted that while all these sources are formally 500 risers, with F500$>$F350$>$F250, the F500/F350 ratio for some is close to 1, so these sources are not as red as some of the 500 risers in the litarature. On this basis they might be expected to lie at somewhat lower redshifts, close to $z\sim 3$ rather than $z > 4$. These five sources represent a complete flux limited sample of 500 risers, ie. it includes all 500 risers in this region, allowing reliable statistical analysis compared to other, heterogeneously selected samples that lack complete followup, or multiwavelength cross identification.

\subsection{Available Multifrequency Data}

The part of the {\em Herschel} XMMLSS field within which our 500 risers were selected is covered by the SERVS {\em Spitzer} survey \citep{m12} which reaches 5$\sigma$ depths of $\sim$2$\mu$Jy in these bands, sufficient to detect an $L\star$ galaxy out to $z\sim$5. \citet{n17} then used the SERVS data as the basis for a multiwavelength catalog with point spread function matched photometry extracted using {\em The Tractor} \citep{l16a,l16b} applied to near-IR data from the VISTA VIDEO survey \citep{j13} and the CFHT Legacy Survey Deep Field 1 (CFHTLS D1) survey at optical wavelengths. This provides a catalog with 11 band optical-to-near-IR psf matched photometry including the $u', g', r', i', z', Z, Y, J, H, K_{s}$ 3.6 and 4.5 $\mu$m bands. More recently, this dataset was re-analysed by \citet{z22} with the addition of GALEX FUV and NUV bands \citep{m05}, HyperSuprimeCam {\em g, r, i, z, y} bands \citep{a18}, SWIRE 5.8, 8, 24, 70 and 160 $\mu$m bands \citep{l03}, and {\em Herschel} bands, though our specific targets lack GALEX, and SWIRE 70 and 160$\mu$m fluxes, and PACS 100 and 160$\mu$m fluxes. The fluxes available for our sources are discussed in more detail later in this paper. \citet{z22} then use CIGALE to fit SED models to all sources in their catalog which combines not only the XMMLSS field but two others which between them will become the Vera C. Rubin Observatory's deep drilling fields to produce a compendium of fluxes and derived properties for over 2 million galaxies. Where available, we use these CIGALE SED fits for our sources later in this paper.

\label{sect:multiwavelength}

\section{SMA Observations}

\subsection{Observations and Data Reduction}

The SMA observations were taken between 2019 Aug 19 - 2019 Sep 3 as part of the SMA program 2019A-S004 (PI: D. Clements). The atmospheric opacity ranged from 0.07 to 0.3, with average values of around 0.1. The SMA uses two receivers, each with two sidebands
of 12 GHz separated by a gap of 8 GHz. We tuned these two receivers to LOs of 213 GHz and 221 GHz respectively to produce 30 GHz of continuous bandwidth spanning 197 GHz to 237 GHz. We used the SMA in COM configuration with 7 or 8 antennae, resulting in a synthesized beam of just over 3 arcseconds. Initial data calibration was performed using the standard IDL-based SMA data reduction package MIR\footnote{https://github.com/qi-molecules/sma-mir}. Firstly, we manually inspected the data, flagging any regions that contained phase jumps and fixing any channels that contained significant noise spikes, before applying the system temperature correction. The passband calibration was completed using the sources 3c454.3, while for flux calibration we used either Neptune or Titan. Gain calibration was carried out by periodically observing the nearby bright quasars 0224+069 and 0241-082. These calibrated data were exported and final imaging and analysis was completed using the MIRIAD software package.

\begin{table}
\begin{tabular}{cccc} \hline
Target& Date&  RMS&Notes\\ \hline
J021856&2019-09-03&0.25mJy&AKA XMM\_M5\\
J022425&2019-09-02&0.23mJy\\
J022427&2019-08-25&0.26mJy\\
J022448& 2019-08-15&0.21mJy\\
J022610& 2019-08-21&0.23 mJy\\ \hline
\end{tabular}
\caption{Observation details}
\end{table}

\begin{figure*}
\begin{tabular}{ccc}
\includegraphics[width=5cm]{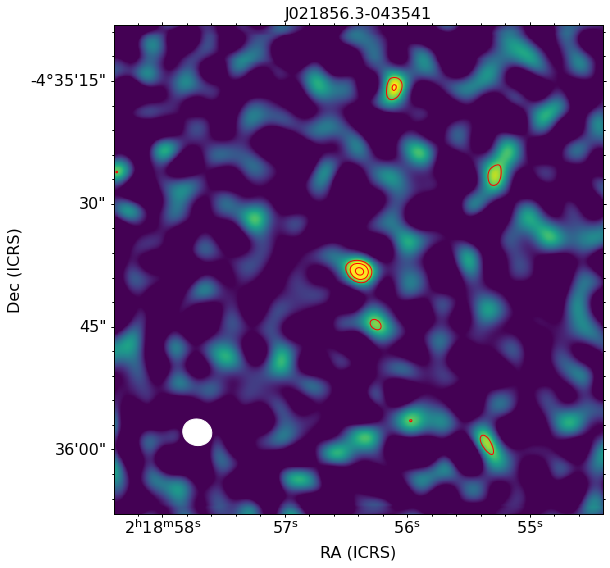}&
\includegraphics[width=5cm]{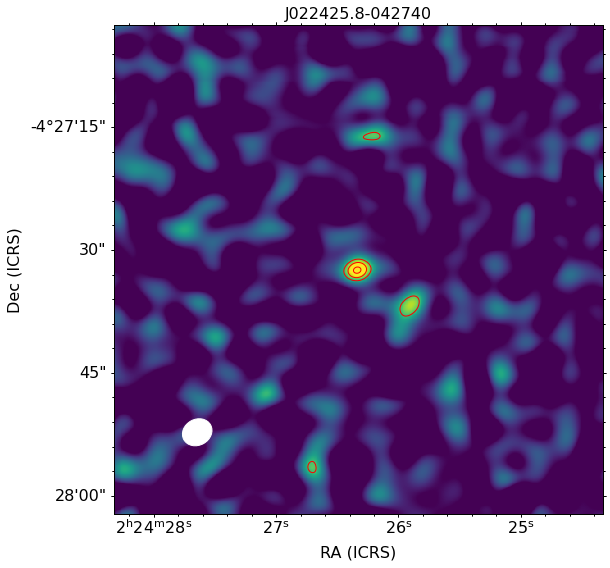}&
\includegraphics[width=5cm]{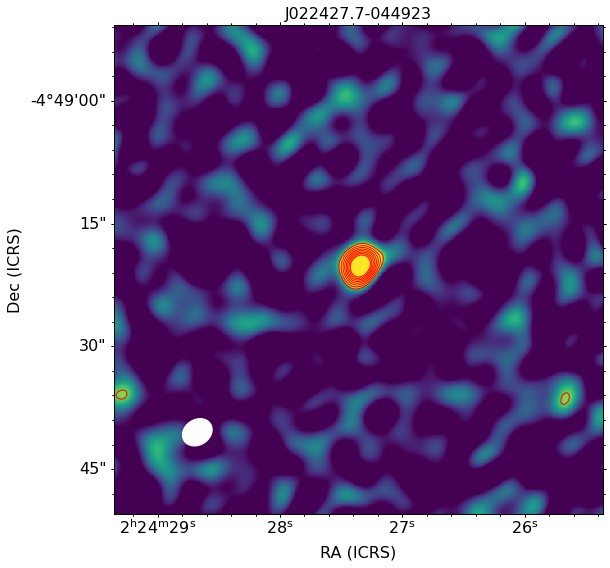}\\
\includegraphics[width=5cm]{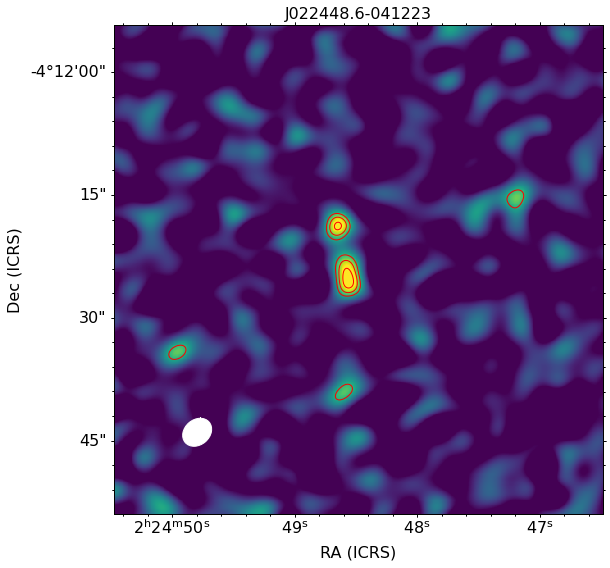}&
\includegraphics[width=5cm]{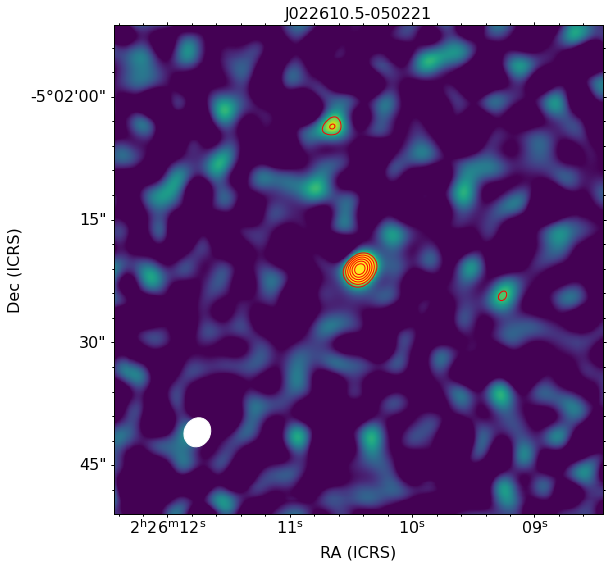}\\
\end{tabular}
\caption{SMA Images of SERVS-XMM 500 Risers. Images are 1 arcminiute on a side. Contours are shown in red and are in increments of 1$\sigma$ starting at 3$\sigma$ and rising from there. The synthesised beam size is just over 3 arcseconds. The beam is shown as a white ellipse in the bottom left of each image.}
\label{fig:sma_images}
\end{figure*}

\begin{table*}
\begin{tabular}{crrrcl}\hline
Source&RA&DEC&200 GHz Flux (mJy)&S/N&Comments\\ \hline
\multicolumn{6}{c}{Secure Detections}\\ \hline
J021856\_1&02:18:56.5&-04:35:38.25&1.3 $\pm$ 0.25&5.2&Centred on Herschel position\\
J021856\_2&02:18:56.1&-04:35:15.9&1.0 $\pm$ 0.25&4.0&Distant from Herschel position, possible 250$\mu$m detection\\
J022425\_1&02:24:26.3&-04:27:32.5&1.2$\pm$0.23&5.2&Centred on Herschel position\\
J022425\_2&02:24:25.9&-04:27:36.8&0.9$\pm$0.23&4.0&Centred on Herschel position\\
J022427\_1&02:24:27.3&-04:49:20.3&3.7$\pm$0.26&14.2&Centred on Herschel position\\
J022448\_1&02:24:48.6&-04:12:25.4&1.2$\pm$0.21&5.7&Centred on Herschel position; Southenmost of likely triple\\
J022448\_2&02:24:48.6&-04:12:19&1.1$\pm$0.21&5.2&Centred on Herschel position; Northenmost of likely triple\\
J022448\_3&02:24:48.6&-04:12:24.1&1.1$\pm$0.21&5.2&Centred on Herschel position; Middle of likely triple\\
J022610\_1&02:26:10.4&-05:02:21&2.0$\pm$0.23&8.5&Centred on Herschel position\\
J022610\_2&02:26:10.7&-05:02:03&0.94$\pm$0.23&4.1&Offset from Herschel position\\ \hline
\multicolumn{6}{c}{Additional Sources}\\ \hline
J021856\_A1&02:18:55.3&-04:35:26.6&0.9$\pm$0.35&3.6&Distant from Herschel position\\
J022427\_A1&02:24:29.85&-04:49:06.7&1.0$\pm$0.26&3.8&Distant from Herschel position\\ \hline
\end{tabular}
\caption{Sources detected by our SMA observations. Where there are multiple sources detected the brightest is given the designation \_1 with fainter sources given \_2 etc. respectively in order of decreasing flux. Additional sources are detected between 3 and 4$\sigma$ significance but have optical/IR counterparts suggesting that they may be genuine submm sources. See text for details.}
\label{table:sma_sources}
\end{table*}

\subsection{SMA Imaging Results}

The results of our SMA observations are shown in Figure \ref{fig:sma_images}, and the positions and derived fluxes of all sources detected at $>4\sigma$ in these images can be found in Table \ref{table:sma_sources}. In all cases at least one source is strongly detected with significance $\gg 4\sigma$ in our final maps. In four cases more than one source is detected in our SMA maps with significance $\geq 4\sigma$ suggesting that these are genuine multiple sources. This result is consistent with \citet{g20} who find that 60\% of F500 $<$ 60 mJy sources are, or are likely to be, multiple on the basis of both detected multiples and undetected sources that are deemed to probably be multiple. We note that one of our multiple sources, J021856, which consists of one 5.3$\sigma$ detected source and a second 4.1$\sigma$ source, was one of the undetected sources in \citet{g20} named XMM\_M5 in that paper, which provides support to the suggestion that the nine undetected F500 $<$ 60 mJy sources in that paper are indeed multiples. See \citet{c23} for further observations of the undetected \citet{g20} sources. Other potential submm sources detected at $3 - 4\sigma$ in these images might also be real detections, increasing the multiplicity of these sources. These are discussed later in this paper in light of potential optical and near-IR support for these lower $\sigma$ sources. They are included in Table \ref{table:sma_sources} as additional sources.

\begin{figure*}
\begin{tabular}{ccc}
 \includegraphics[width=5cm]{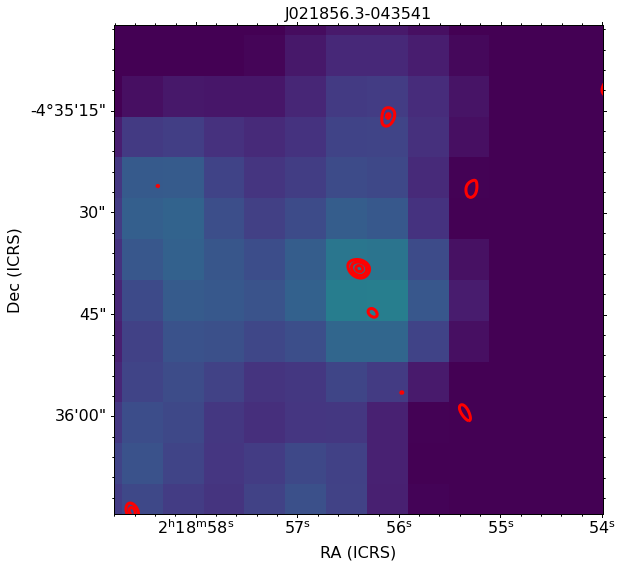}&
  \includegraphics[width=5cm]{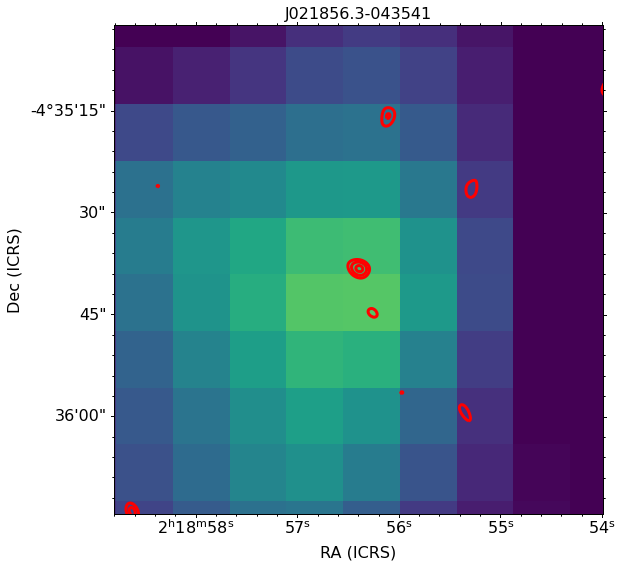}&
   \includegraphics[width=5cm]{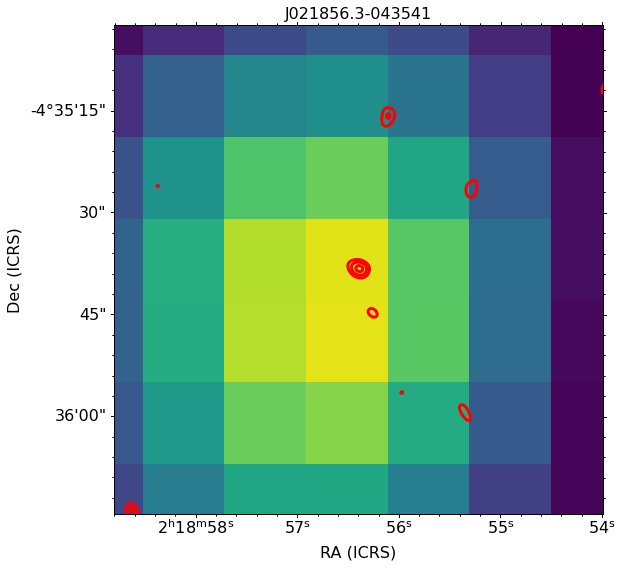}\\   
   
         \includegraphics[width=5cm]{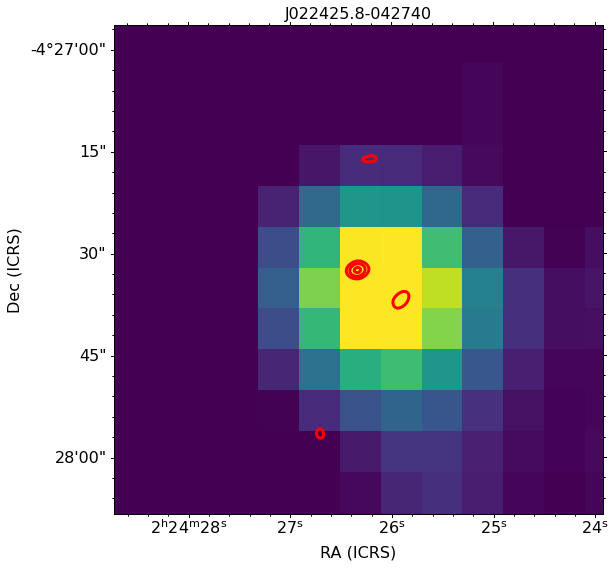}&
  \includegraphics[width=5cm]{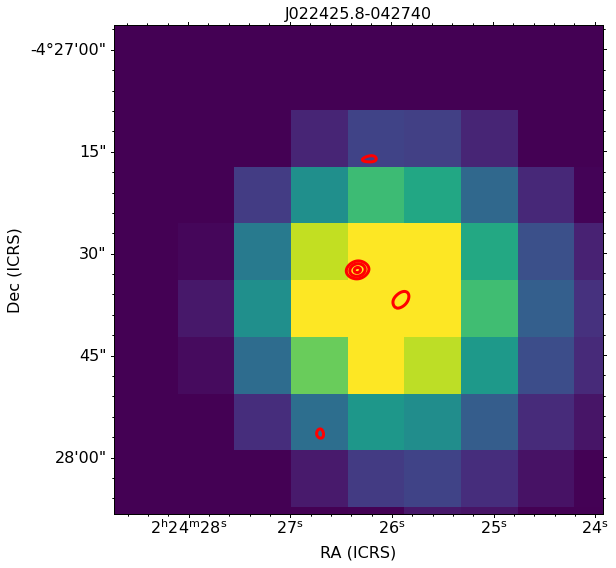}&
   \includegraphics[width=5cm]{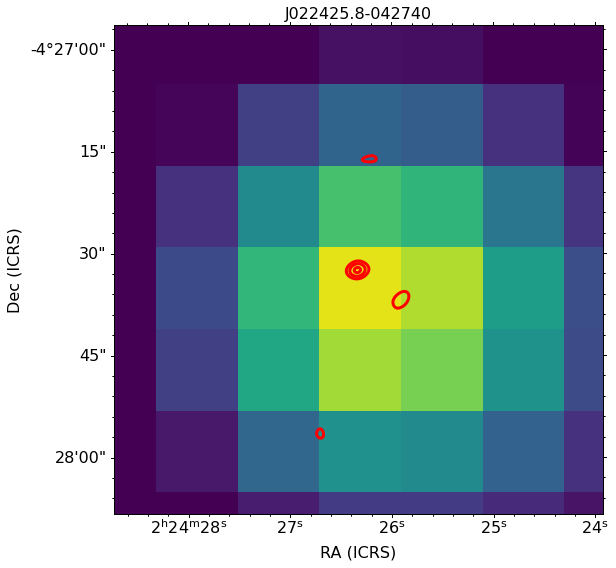}\\ 
   
    \includegraphics[width=5cm]{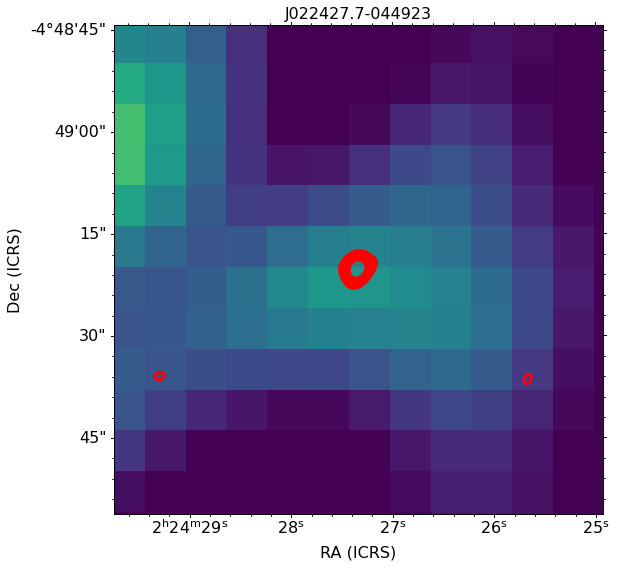}&
  \includegraphics[width=5cm]{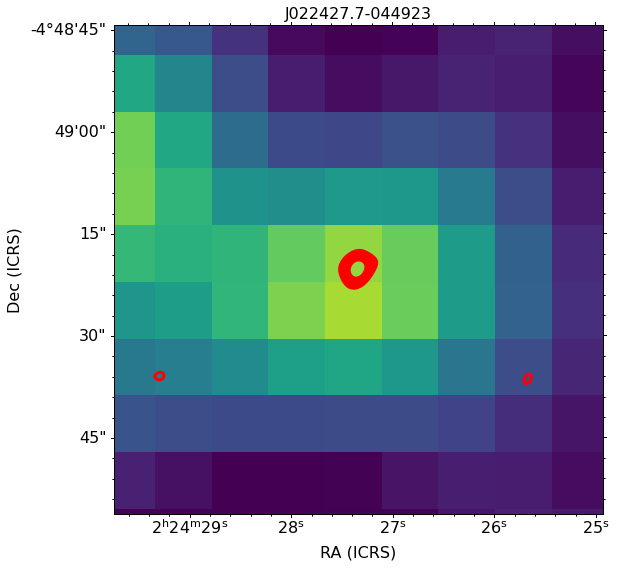}&
   \includegraphics[width=5cm]{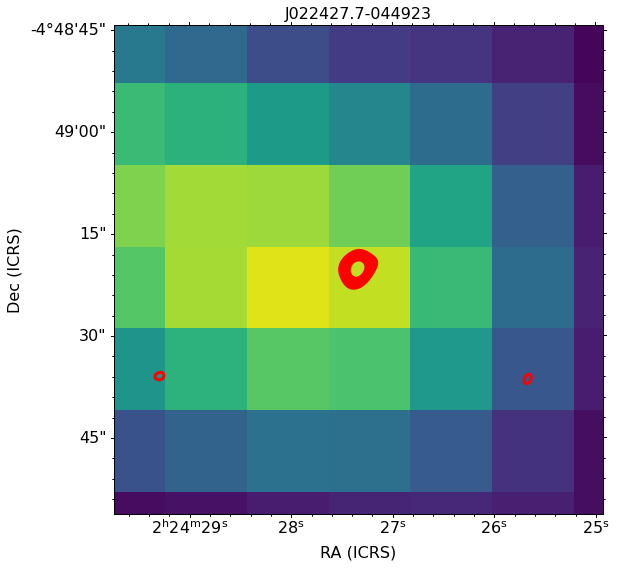}\\  
   
       \includegraphics[width=5cm]{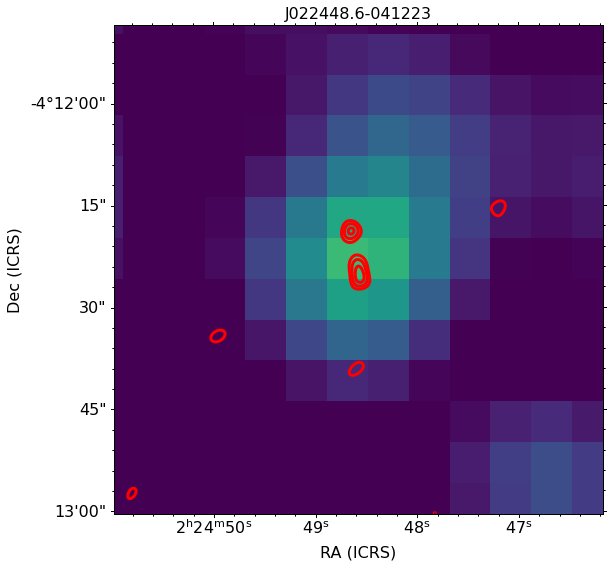}&
  \includegraphics[width=5cm]{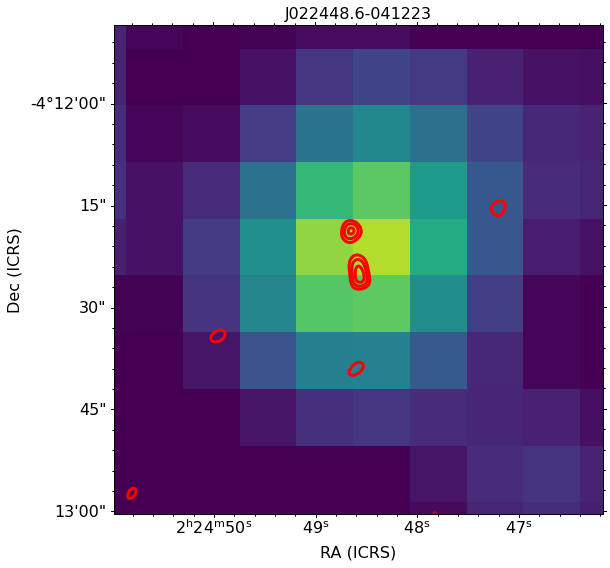}&
   \includegraphics[width=5cm]{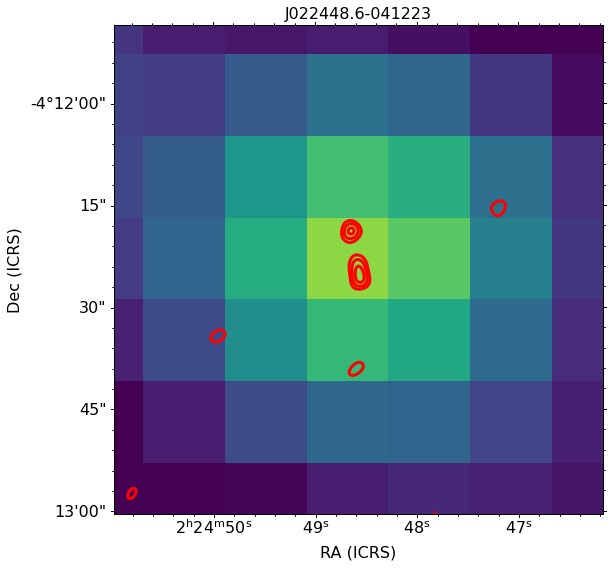}\\ 
   
       \includegraphics[width=5cm]{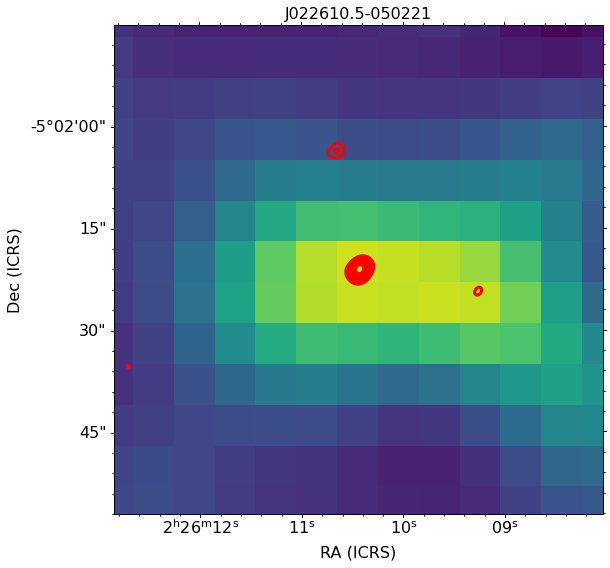}&
  \includegraphics[width=5cm]{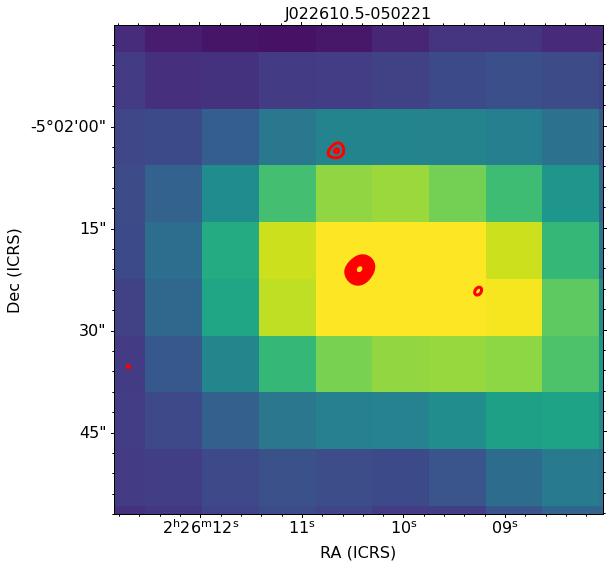}&
   \includegraphics[width=5cm]{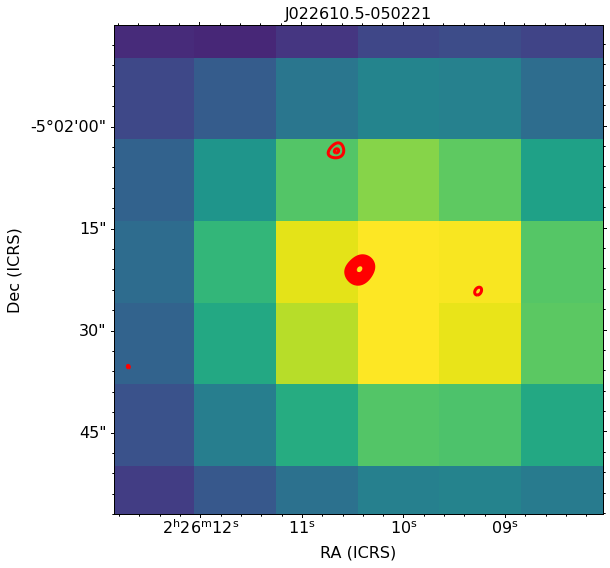}\\  

\end{tabular}
\caption{Herschel images of the SERVS-XMM 500 risers. Images are at 250, 350 and 500$\mu$m from left to right, from Herschel, with contours starting at 3$\sigma$ and rising in 1$\sigma$ intervals from our SMA observations. Images are 72 arcseconds on a side.}
\label{fig:sma_herschel_images}
\end{figure*}

In Figure \ref{fig:sma_herschel_images} we show our SMA contours overlaid on the Herschel images of these sources at 250, 350 and 500$\mu$m. In all cases the brightest SMA sources are located at positions matching the Herschel sources. This demonstrates that our SMA data is appropriate for the key goal of this paper, to find SMA counterparts to the Herschel sources at higher resolution and thence obtain clear optical/NIR identifications for these 500 risers. A detailed discussion of the submm properties of each source can be found below alongside the results of the optical/NIR cross identification process.

\section{Optical/NIR Identifications}

The main goal of the current work is to use the arcsecond accuracy positions from the SMA to obtain cross identifications with the deep multiwavelength optcial/near-IR data available in the XMM survey region. This will allow a much better and more comprehensive understanding of these objects than is possible from the Herschel and SMA data alone. The optical/NIR data available in this field is described in Section \ref{sect:multiwavelength} and includes deep imaging at U, g, r, i, z, y, J, H, Ks bands and SERVS [3.6] and [4.5] micron imaging. 

\begin{figure*}
\begin{tabular}{cccc}
\includegraphics[width=4cm]{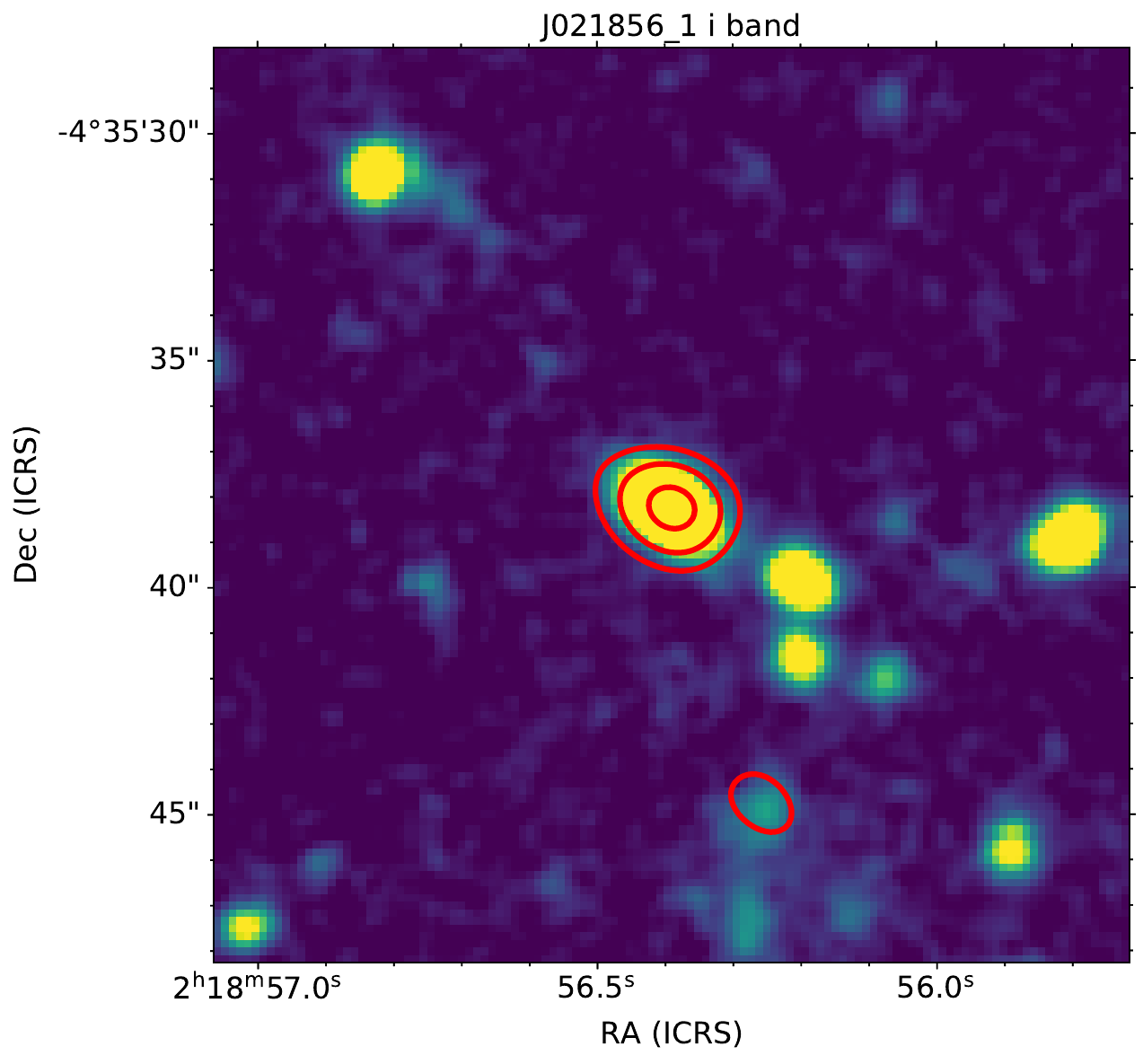}&
\includegraphics[width=4cm]{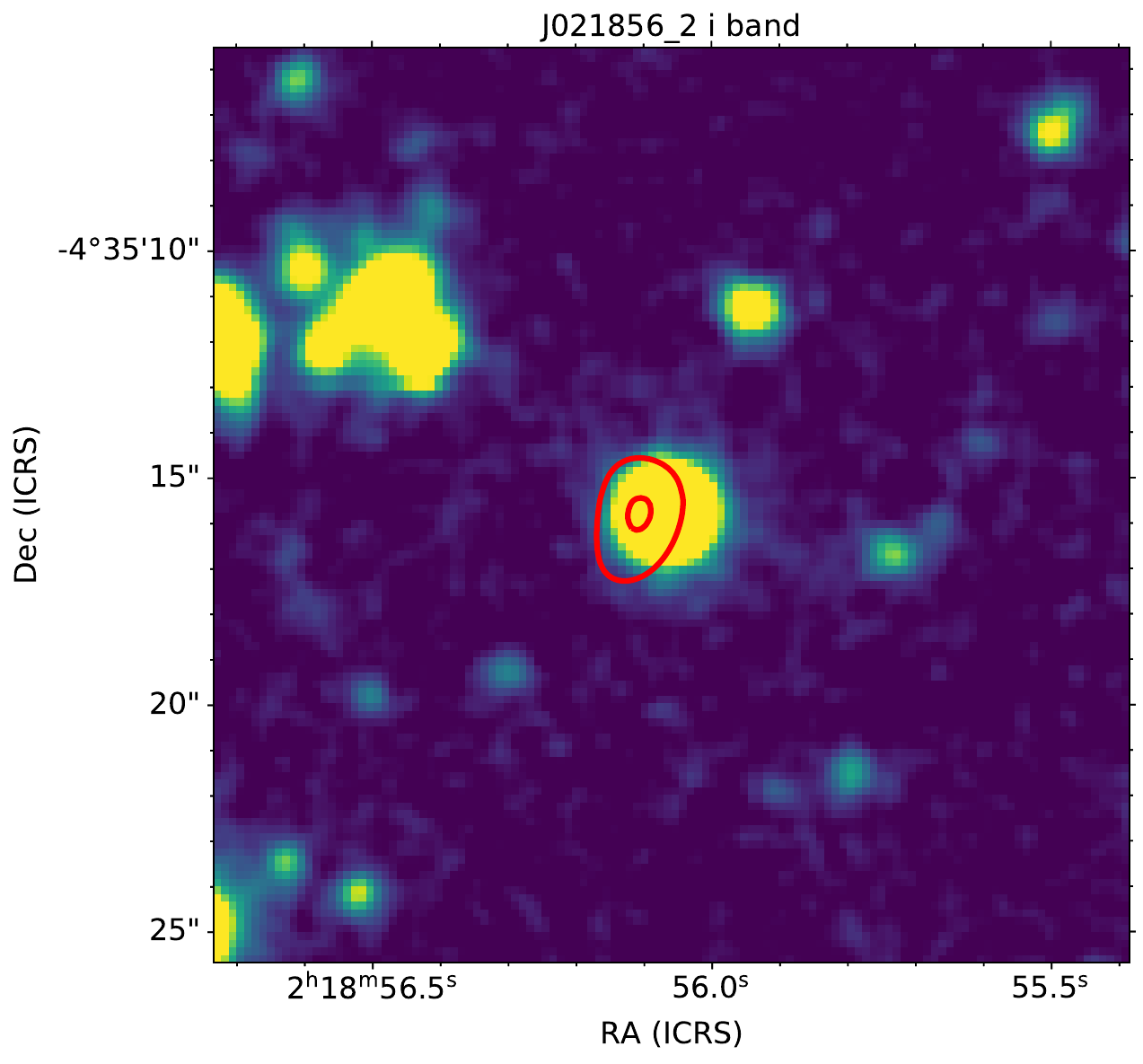}&
\includegraphics[width=4cm]{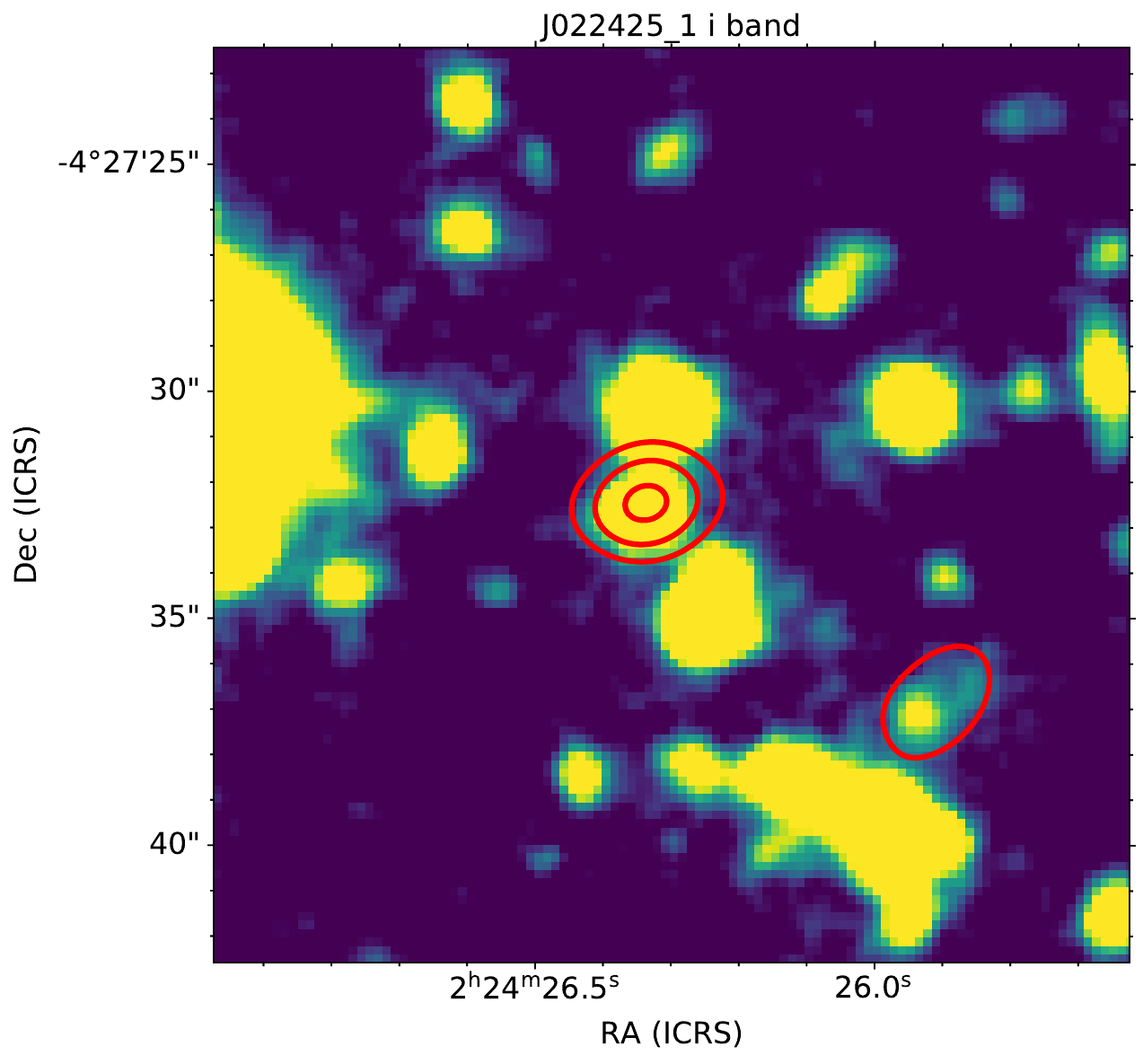}&
\includegraphics[width=4cm]{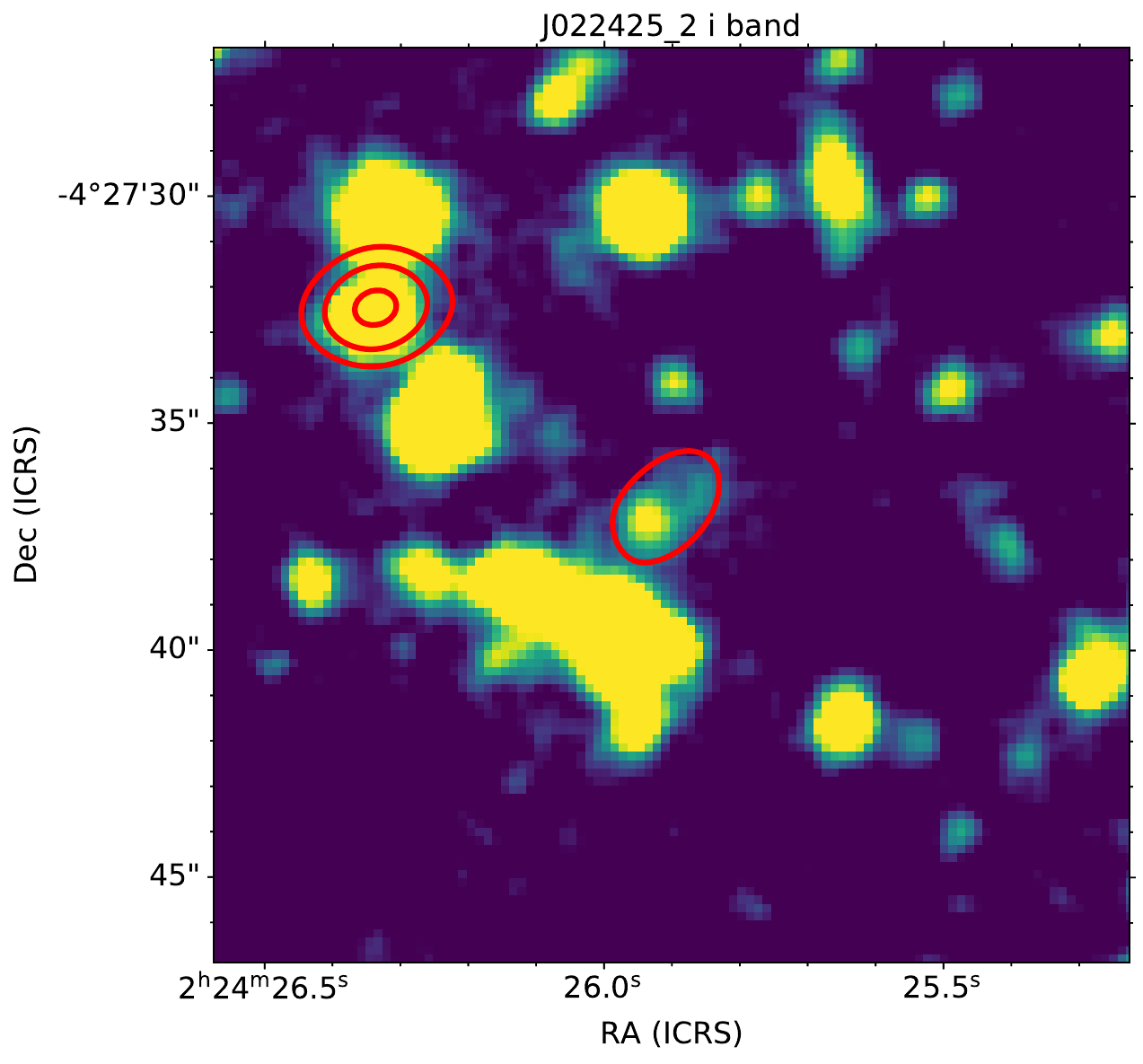}\\
\includegraphics[width=4cm]{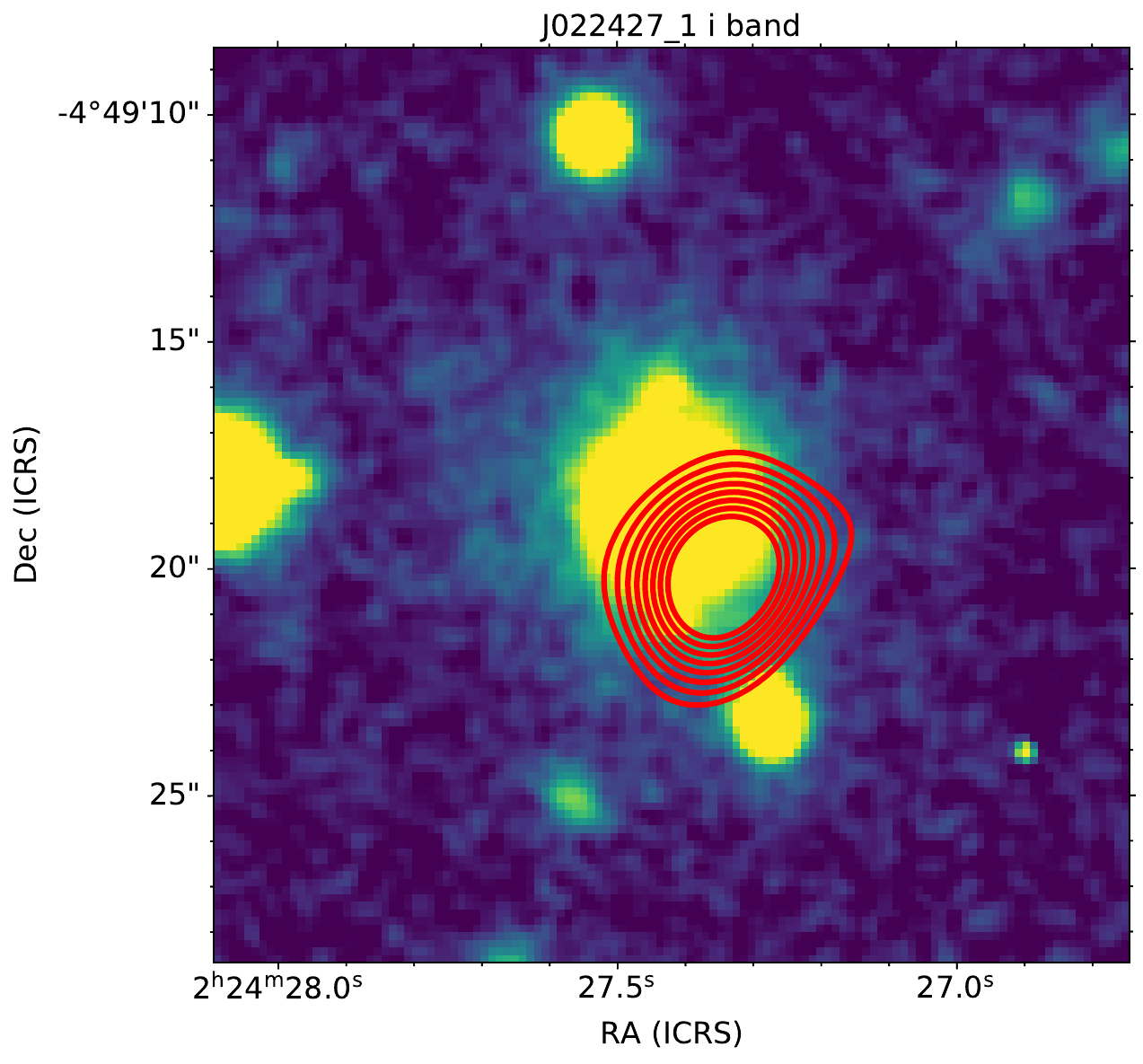}&
\includegraphics[width=4cm]{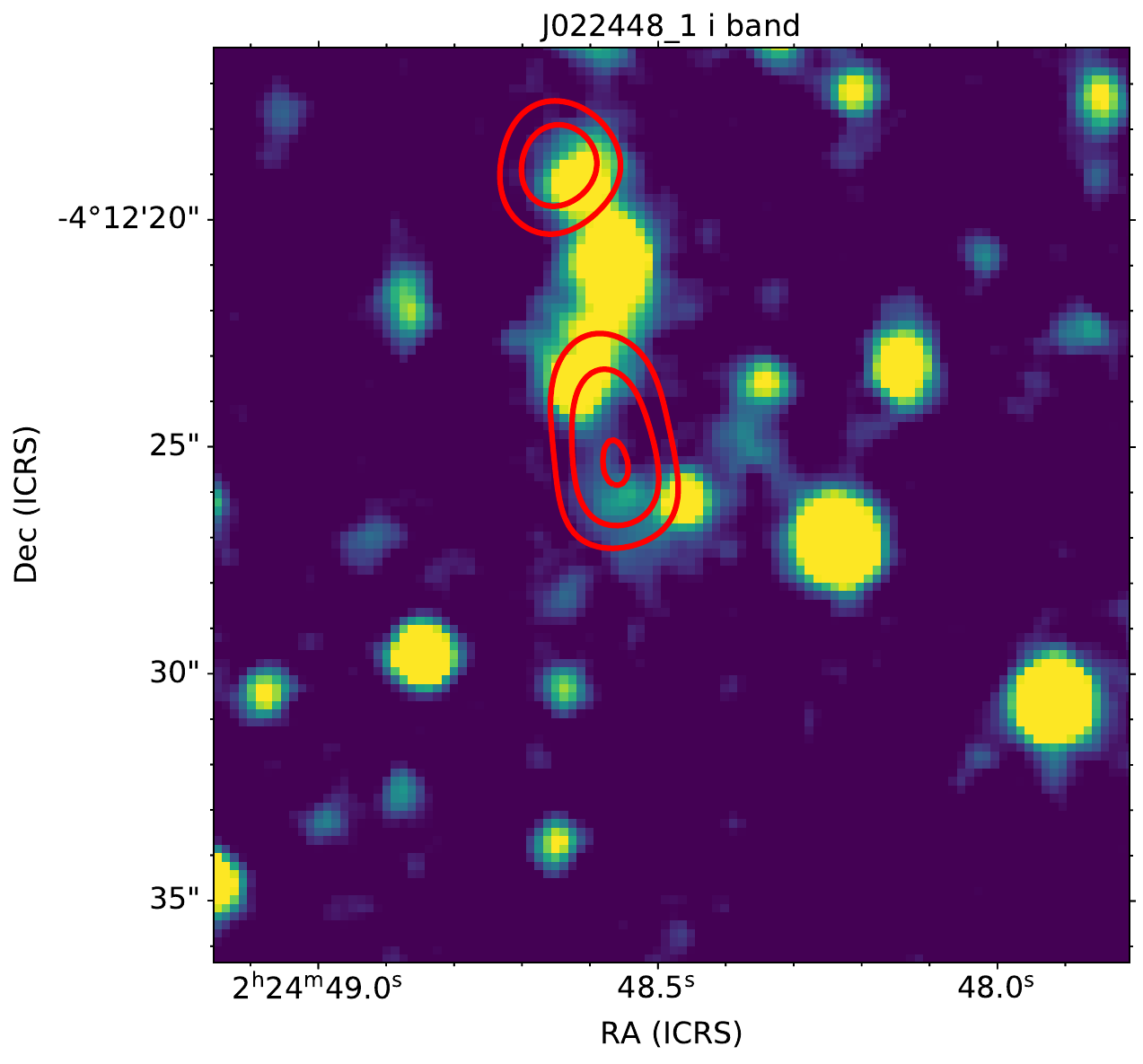}&
\includegraphics[width=4cm]{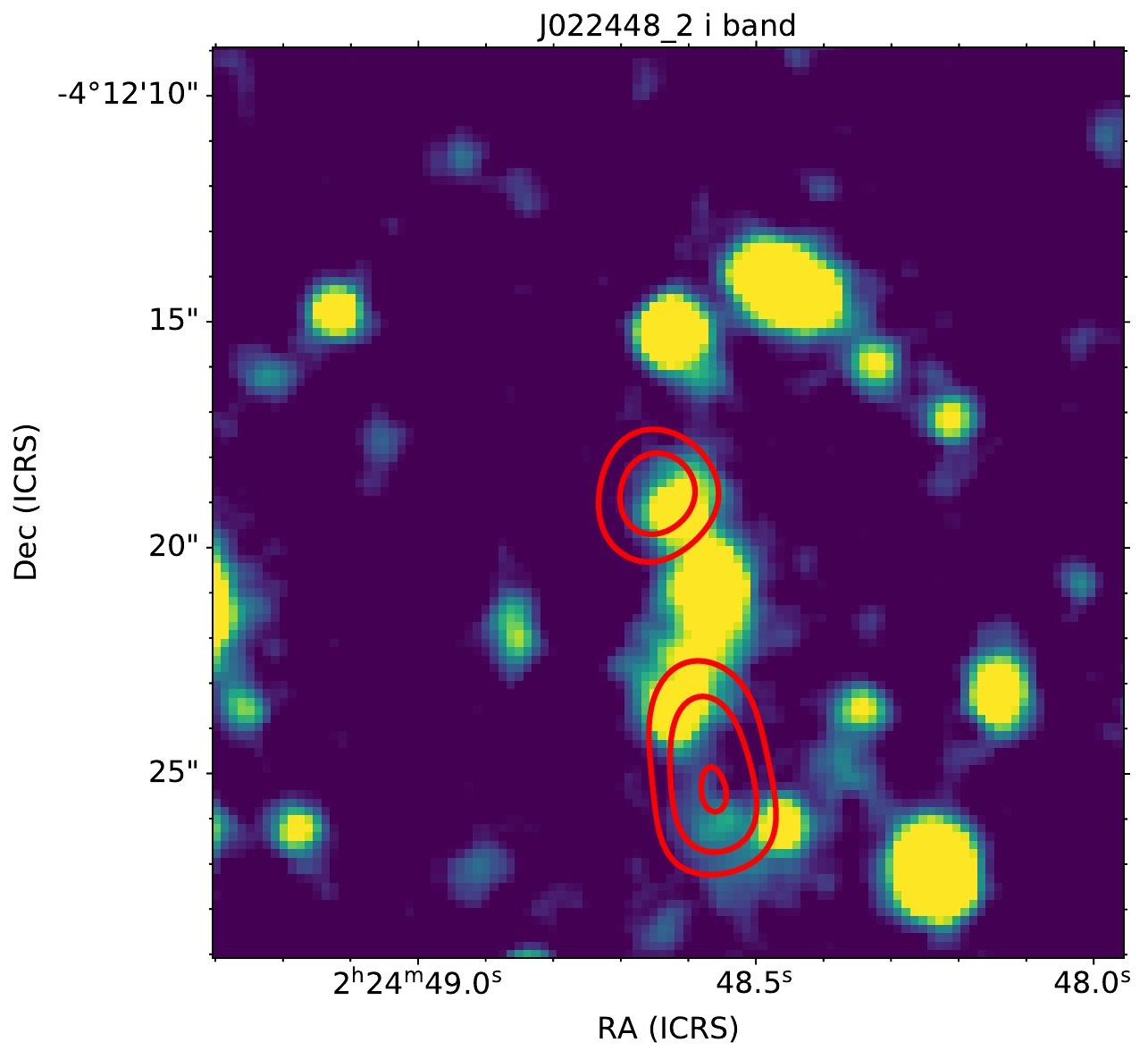}&
\includegraphics[width=4cm]{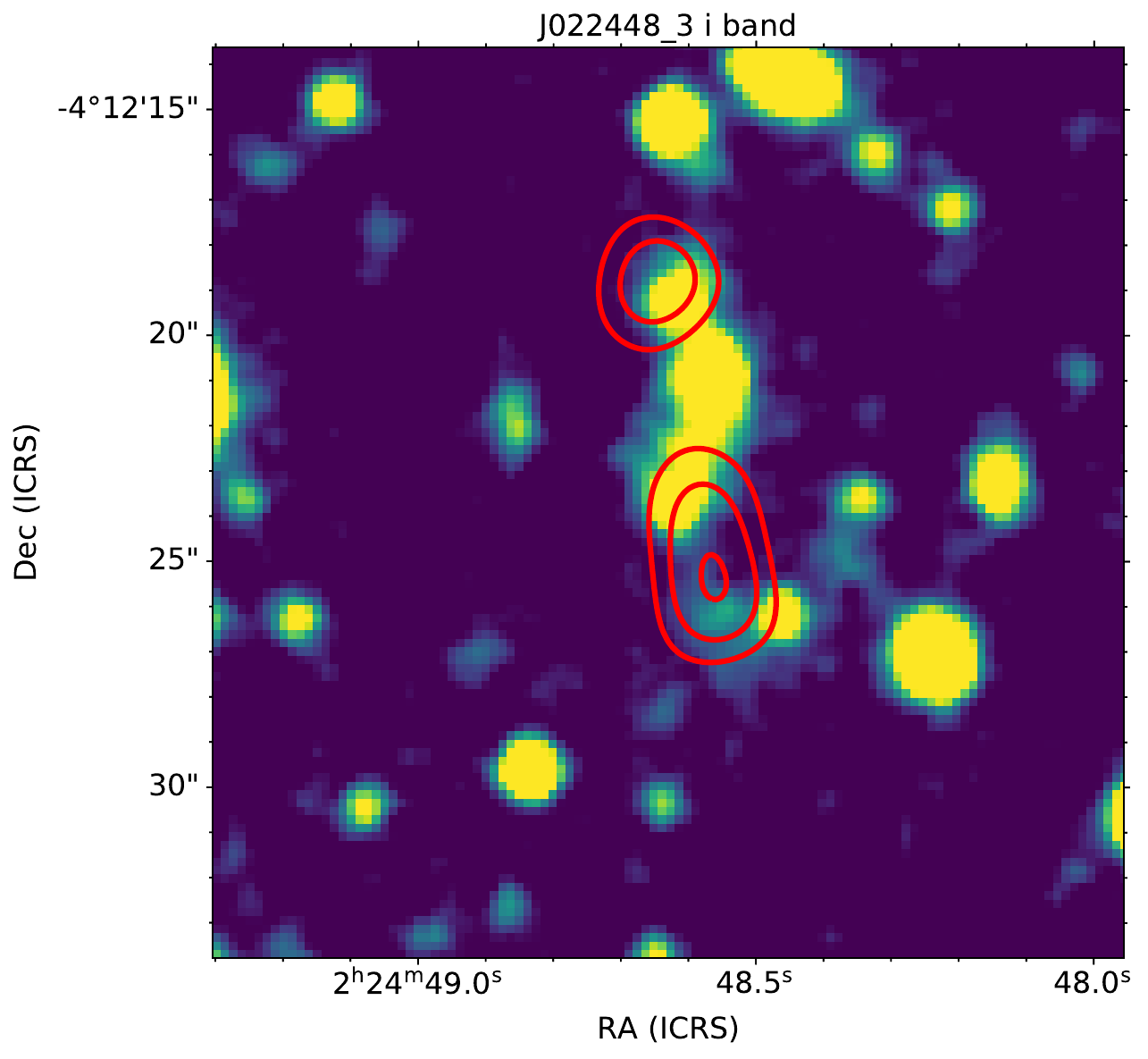}\\
\includegraphics[width=4cm]{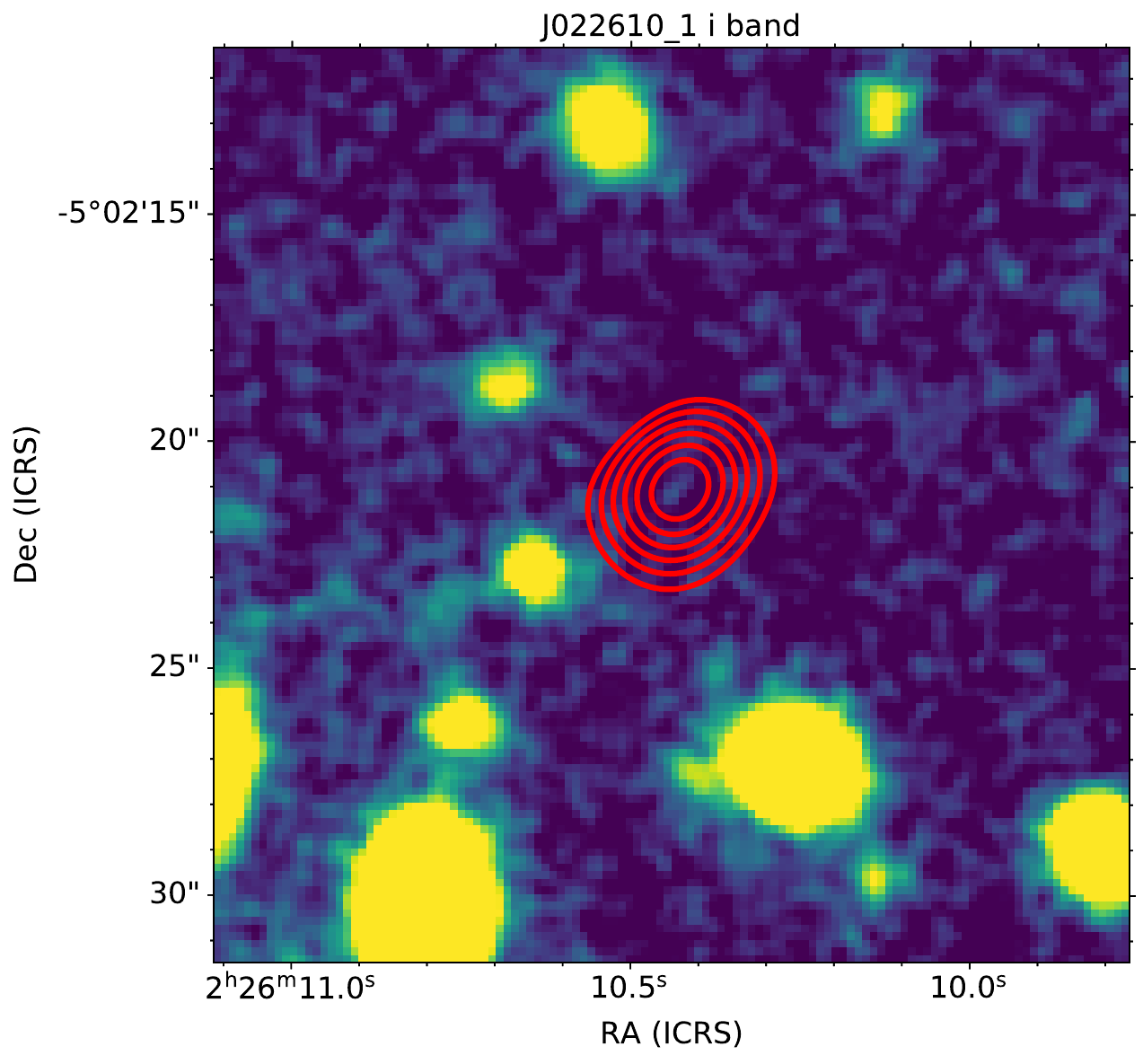}&
\includegraphics[width=4cm]{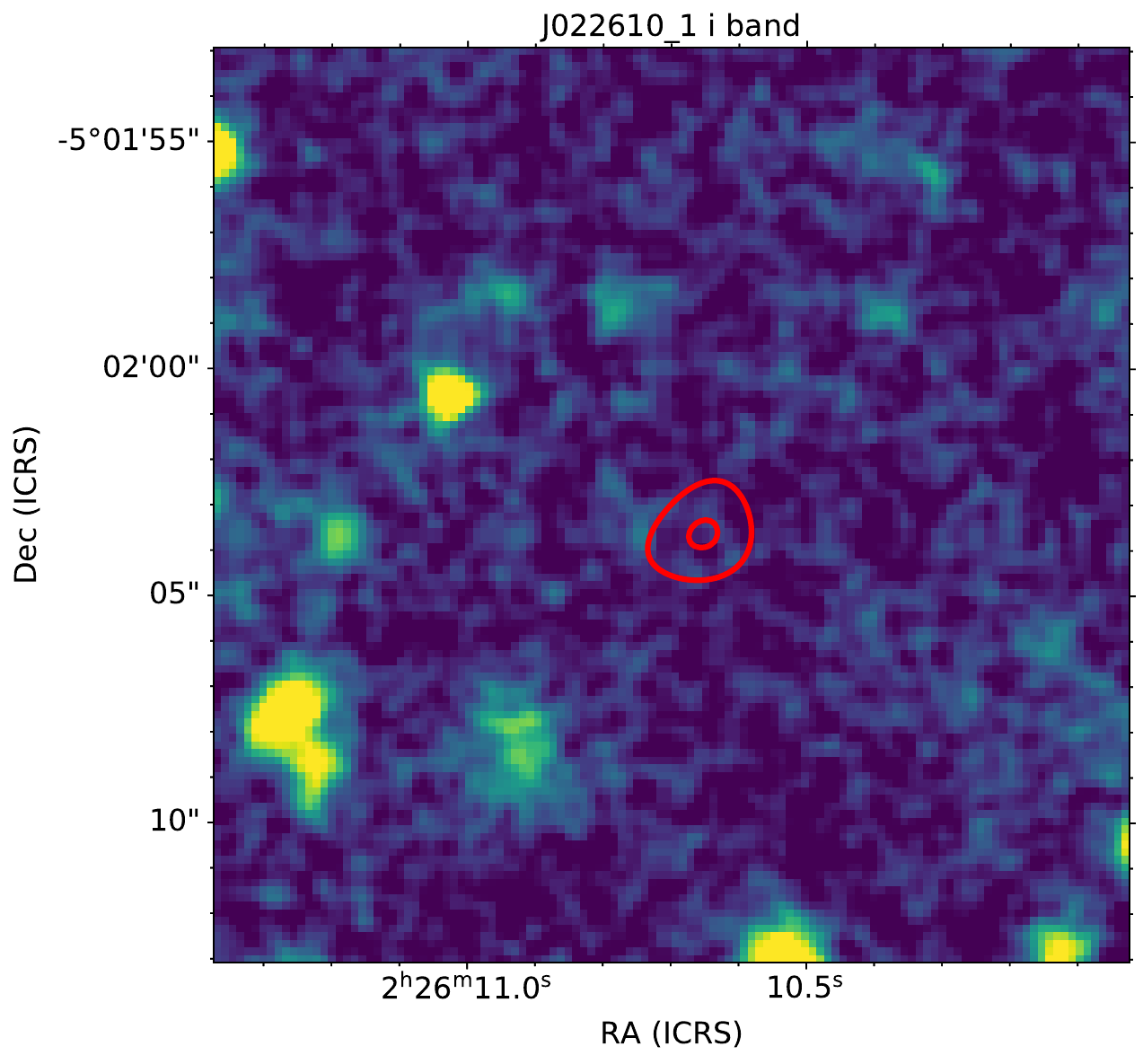}&
\includegraphics[width=4cm]{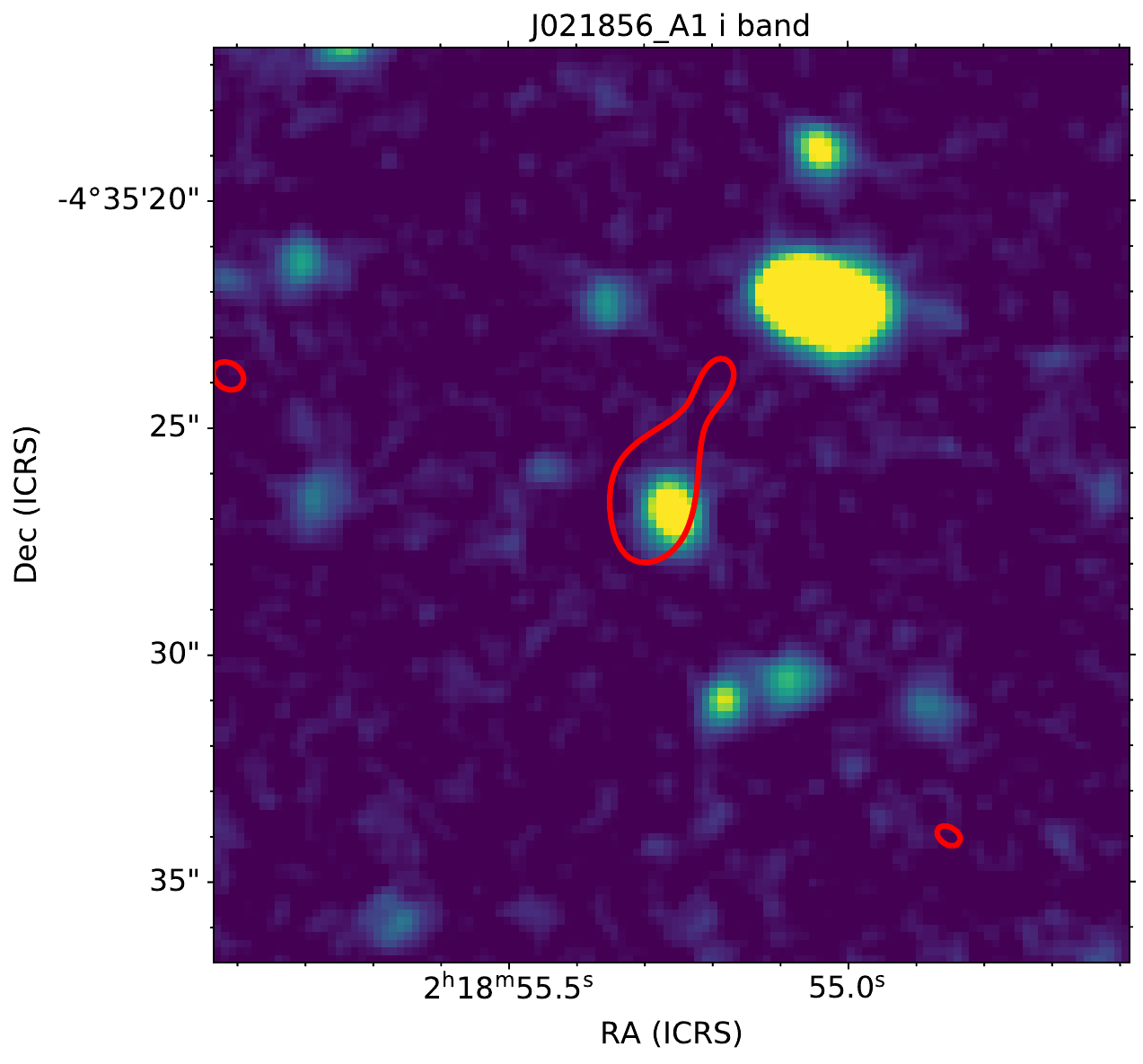}&
\includegraphics[width=4cm]{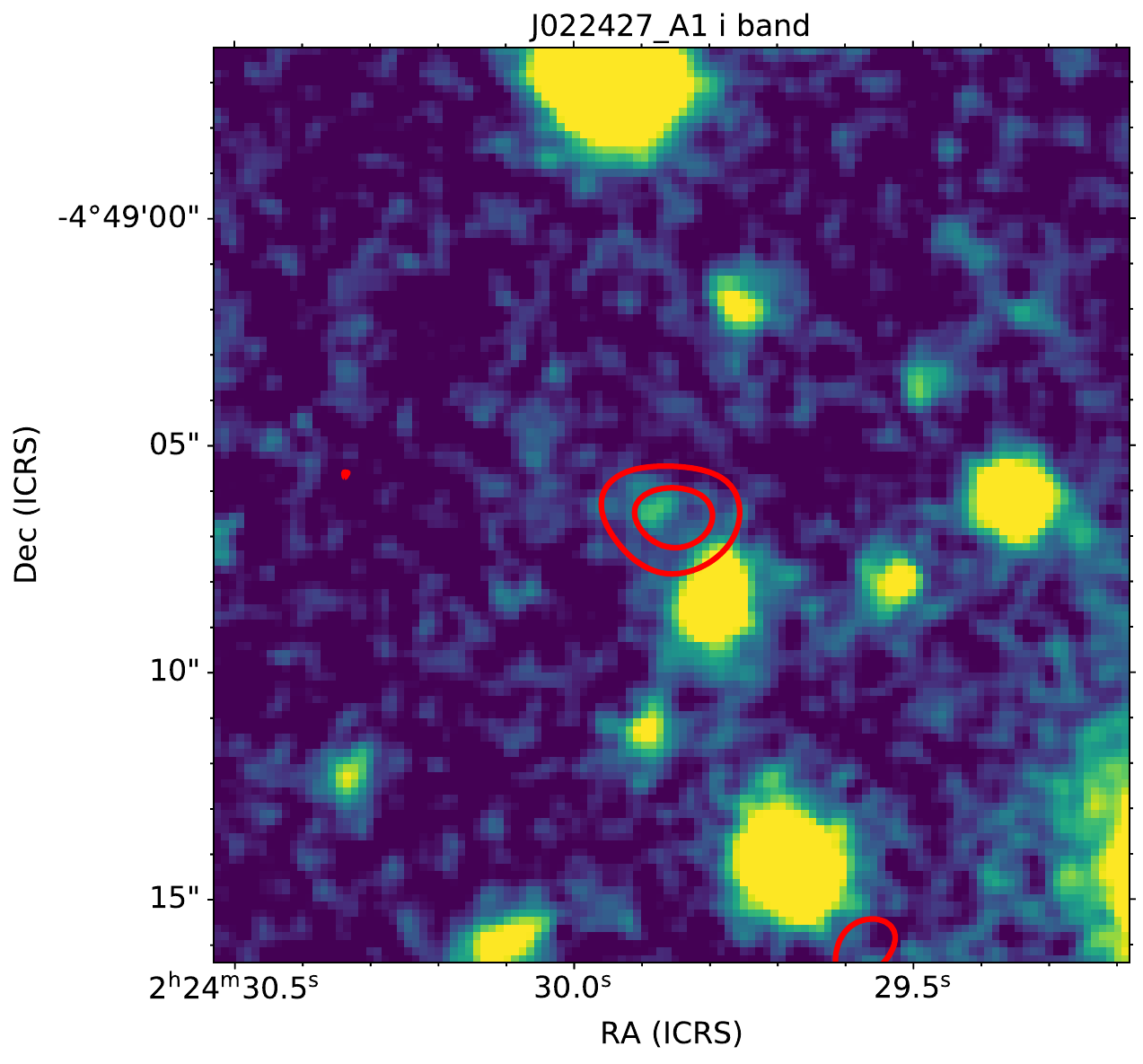}\\
\includegraphics[width=4cm]{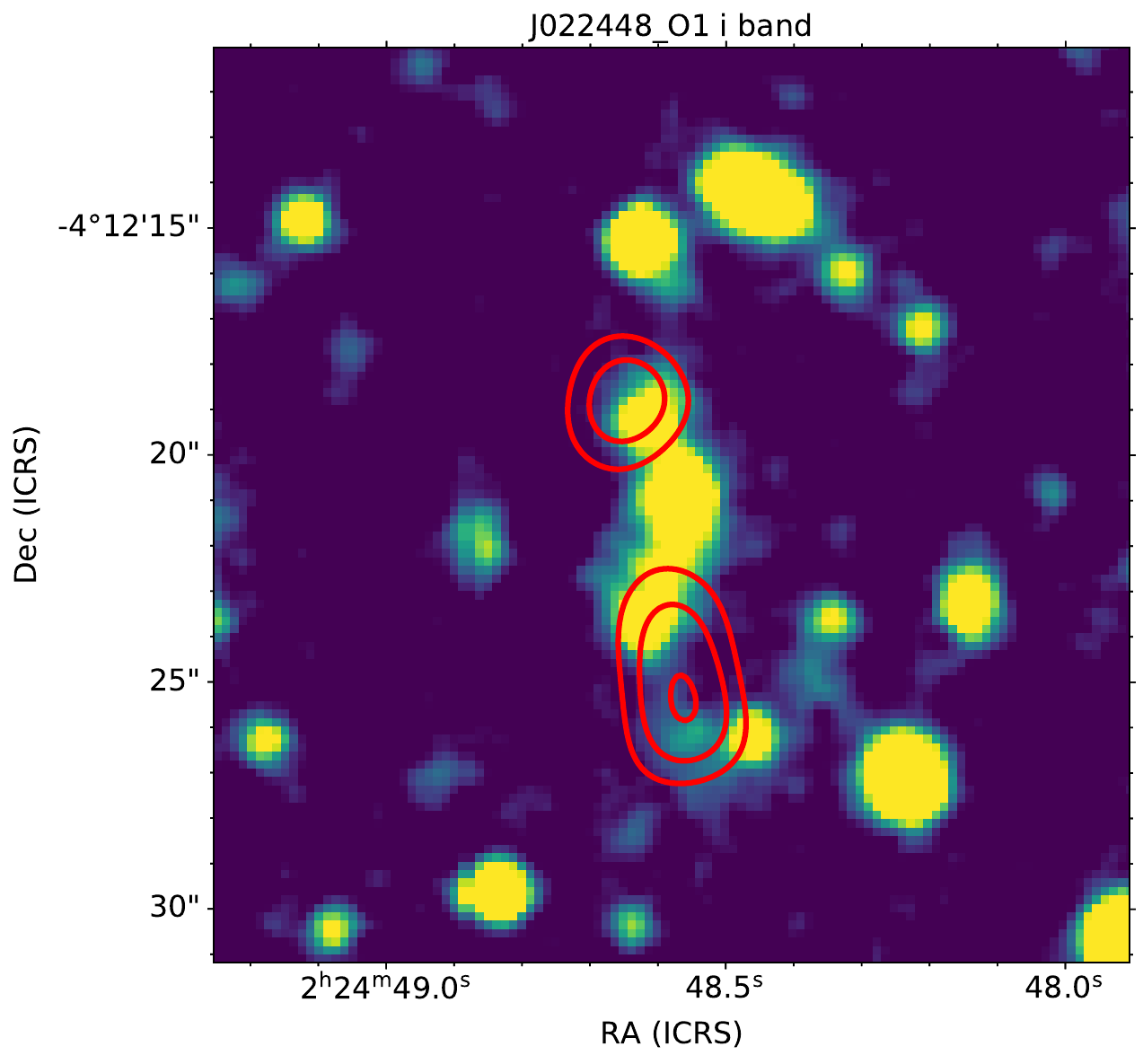}\\
\end{tabular}
\caption{Images of 500 risers. {\em i} band images of our 500 riser sources with the SMA contours overlaid. Also shown are the additional sources. These provide secure optical identifications for nearly all of our sources. Images are from the HSC Public Survey except for the sources associated with 02448 and 022425 which come from the CFHT Lensing Survey \citep{h12}.}
\label{fig:sma_opt}
\end{figure*}

The process we adopt for finding optical/near-IR cross identifications is to overlay the SMA contours on optical/near-IR images as appropriate to identify the cross-ID, and then to extract the psf-matched photometry and astrometry for this source from the TRACTOR generated catalogue in \citet{n17}. Optical images with the SMA contours used in this process overlaid are shown in Figure \ref{fig:sma_opt} and the resulting photometry is given in Table \ref{table:opt_phot}.

Additional fluxes beyond those in the TRACTOR catalog have been found for some sources from the literature. Fluxes at 24$\mu$m come from the SWIRE survey catalog \citep{l03} for those sources that were detected in this band. A 3$\sigma$ upper limit of 150$\mu$Jy at 24$\mu$m, consistent with the sensitivity of the SWIRE survey in the XMM field \citep{s08} is assumed for sources not detected at this wavelength, consistent with inspection of the SWIRE 24$\mu$m images at the positions of these sources. Source J021856\_1 has an additional flux point at 850$\mu$m from \citet{c23} of 8.1$\pm$3.3 mJy. See \citet{c23} for details. These additional fluxes are listed in Table \ref{table:additional}.

This process also allows us to assess the reality of sources detected in the SMA observations below our nominal 4$\sigma$ detection threshold. Sources with $S/N > 3$ in the SMA observations that have an optical counterpart are deemed to be plausibly real. The details of these sources are given in Table \ref{table:sma_sources} with photometry given in Table \ref{table:opt_phot} and images shown in Figure \ref{fig:sma_opt}. These additional sources are given the designation A and are numbered from 1 in order of decreasing submm flux, thus we would have J021856\_A1 and J021856\_A2 if there were two companions to J021856. We also include in this table of additional sources an optical source that appears to be associated with the triple submm sources revealed by our SMA imaging of J022448.

Details for all of these sources are given in the discussion section.

\begin{table*}
\begin{tabular}{ccccccccccc}\hline
Name&J021856\_1&J021856\_2&J022425\_1&J022425\_2&J022427\_1&J022448\_1&J022448\_2&J022448\_3&J022610\_1&J022610\_2\\ \hline
ID&1083202&1085742& 269071&430002&160912&326636&425131&333243&448228&No ID\\
RA&2:18:56.39&2:18:56.06&2:24:26.35&2:24:25.9&2:24:27.42&2:24:48.48&2:24:48.63&2:24:48.63&2:26:10.44&\\
DEC&-4:35:38.2&-4:35:15.6&-4:27:32.36&-4:27:36.9&-4:49:18.6&-4:12:26.3&-4:12:19.0&-4:12:23.7&-5:02:21.4&\\
U& 0.96$\pm$0.16&0.57$\pm$0.47&0.06$\pm$0.04&0.12$\pm$0.04&0.76$\pm$0.10&0.10$\pm$0.05&0.03$\pm$0.05&0.02$\pm$0.04&&\\
g&1.6 $\pm$0.15&4.78$\pm$0.43&0.23$\pm$0.06&0.14$\pm$0.05&4.97$\pm$0.46&0.30$\pm$0.08&0.26$\pm$0.08&0.24$\pm$0.06&&\\
r&3.60$\pm$0.33&11.47$\pm$1.43&0.81$\pm$0.10&0.18$\pm$0.07&23.7$\pm$2.1&0.42$\pm$0.11&0.44$\pm$0.10&0.57$\pm$0.09&&\\
i&4.55$\pm$0.42&12.75$\pm$1.14&1.11$\pm$0.15&0.24$\pm$0.10&48.4$\pm$4.4&0.42$\pm$0.15&0.49$\pm$0.15&0.52$\pm$0.11&&\\
z&5.49$\pm$0.51&18.32$\pm$1.65&1.66$\pm$0.27&0.25$\pm$0.18&64.5$\pm$5.8&0.53$\pm$0.32&0.53$\pm$0.31&0.54$\pm$0.23&&\\
y&6.27$\pm$0.60&18.10$\pm$1.62&2.1$\pm$0.5&0.38$\pm$0.40&77.9$\pm$7.1&0.61$\pm$0.62&0.54$\pm$0.58&0.61$\pm$0.43&&\\
Z&5.38$\pm$0.51&18.55$\pm$1.67&2.46$\pm$0.25&0.36$\pm$0.11&65.7$\pm$5.9&0.69$\pm$0.17&0.70$\pm$0.17&0.72$\pm$0.14&&\\
Y&6.6$\pm$0.60&18.24$\pm$1.64&3.58$\pm$0.33&0.69$\pm$0.10&84.1$\pm$7.6&0.74$\pm$0.11&0.98$\pm$0.13&1.08$\pm$0.12&&\\
J&6.25$\pm$0.62&21.3$\pm$1.9&5.46$\pm$0.53&1.62$\pm$0.21&99.4$\pm$9.0&1.25$\pm$0.25&1.41$\pm$0.25&1.20$\pm$0.20&&\\
H&8.27$\pm$0.82&23.3$\pm$2.1&11.6$\pm$1.1&3.77$\pm$0.41&155.4$\pm$14.0&2.57$\pm$0.43&3.83$\pm$0.49&3.26$\pm$0.40&&\\
Ks&11.82$\pm$1.20&25.6$\pm$2.3&22.9$\pm$2.1&8.70$\pm$0.85&199.5$\pm$18.0&6.38$\pm$0.8&6.57$\pm$0.77&4.95$\pm$0.61&1.02$\pm$0.25&\\
3.6&12.50$\pm$0.17&23.28$\pm$0.19&41.23$\pm$0.54&31.40$\pm$0.35&116.4$\pm$0.4&13.8$\pm$0.1&16.6$\pm$0.2&12.0$\pm$0.2&6.49$\pm$0.39&\\
4.5&16.0$\pm$0.23&22.27$\pm$0.19&48.9$\pm$0.6&44.62$\pm$0.47&97.3$\pm$0.4&20.9$\pm$0.2&24.6$\pm$0.2&17.1$\pm$0.2&11.7$\pm$0.4&\\
Spec\_$z$&0.633&&&&0.4956&\\
Sep&1.6&0.6&0.8&0.3&2.4&2.0&0.4&0.6&0.7&\\
\hline
\multicolumn{11}{c}{Additional Sources}\\ \hline
Name&J021856\_A1&J022427\_A1&J022448\_O1&\\
ID&1091585&170449&339093&\\
RA&2:18:55.26&2:24:29.87&2:24:48.58&\\
DEC&-4:35:26.7&-4:49:06.3&-4:12:21.1&\\
U&0.32$\pm$0.09&0.02$\pm$0.06&0.08$\pm$0.02&\\
g&0.43$\pm$0.05&0.23$\pm$0.10&0.30$\pm$0.04&\\
r&0.48$\pm$0.05&0.51$\pm$0.15&0.49$\pm$0.05&\\
i&0.68$\pm$0.07&0.66$\pm$0.23&0.33$\pm$0.04&\\
z&1.03$\pm$0.11&0.80$\pm$0.57&0.37$\pm$0.08&\\
y&1.05$\pm$0.17&0.84$\pm$0.86&0.37$\pm$0.13&\\
Z&1.11$\pm$0.14&1.12$\pm$0.19&0.53$\pm$0.08&\\
Y&1.59$\pm$0.15&1.56$\pm$0.18&0.58$\pm$0.06&\\
J&1.21$\pm$0.22&2.19$\pm$0.30&0.49$\pm$0.09&\\
H&1.73$\pm$0.28&4.55$\pm$0.55&0.90$\pm$0.16&\\
Ks&1.87$\pm$0.41&7.10$\pm$0.81&0.96$\pm$0.24&\\
3.6&2.19$\pm$0.15&15.46$\pm$0.33&7.58$\pm$0.16&\\
4.5&1.66$\pm$0.19&25.64$\pm$0.37&9.94$\pm$0.02&\\
Sep&0.5&0.5\\
Notes&&&Associated&\\
&&&optical source&\\
\hline
\end{tabular}
\caption{Optical to near-IR fluxes for 500 risers. All fluxes are in $\mu$Jy. Sep indicates the separation between the SMA position and the optical position from the catalog in arcseconds. Spectroscopic redshifts where available is listed. ID number refers to the SERVS ID in \citet{n17} catalog. Data are from CFHT (U), HSC g (to y), VIDEO (Z to Ks) and SERVS (3.6 and 4.5). See \citet{n17} for details. A blank flux entry indicates non-detection in \citet{n17} TRACTOR analysis. The source J022610\_2 lacks any optical/IR identification in Nyland catalog. None of the associated sources have a spectroscopic redshift.}
\label{table:opt_phot}
\end{table*}

\begin{table*}
\begin{tabular}{ccccc}\hline
Source&5.8$\mu$m&8$\mu$m&24$\mu$m&850$\mu$m\\ \hline
J021856\_1&&&0.37 $\pm$ 0.026 mJy&8.1$\pm$3.3mJy\\
J022425\_1&&&0.78 $\pm$ 0.023 mJy&\\
J022425\_2&&&1.1 $\pm$ 0.023$^*$ mJy\\
J022427\_1&0.050 $\pm$0.004 mJy\\
J022448\_1&&0.030 $\pm$ 0.005 mJy\\ \hline
\end{tabular}
\caption{Additional fluxes for specific sources extracted from the SWIRE survey (5.8, 8 and 24$\mu$m) and from \citet{c22} (850$\mu$m). $^*$ the 24 $\mu$m flux for this source is potentially blended with that of a nearby non-SMA detected source.}
\label{table:additional}
\end{table*}

\section{The Redshifts of 500 Risers}

We now have multiwavelength data from optical to submm for all our 500 riser sources and for several additional sources as well, allowing far greater insights into the nature of these sources than is possible with the Herschel and submm data alone. Our next step is to obtain photometric redshifts for our sources. We use three separate approaches for this: firstly we use the MMPz tool \citep{c20} to determine photometric redshift estimates for the far-IR/submm data only; secondly, where available, we use high quality literature photo-zs based on optical-to-NIR photometry or, in two cases, optical spectroscopic redshifts; thirdly, two of our sources lack sufficient optical-NIR data to be included in previous studies. We therefore use EAzY to derive a photometric redshift for the source where there is limited near-IR data (J022610\_1). For the final source which is an optical-NIR blank field, we are forced to rely on the MMPz redshift estimate alone. Details of these analyses are given below.

\subsection{Far-IR Only Photometric Redshifts}

MMPz\footnote {http://www.as.utexas.edu/~cmcasey/mmpz.html} \citep{c20} is a tool for deriving photometric redshift estimates from far-IR/submm data. It takes account of the intrinsic breadth of rest-frame SEDs and uses fluxes as well as colours to derive the best redshift constraints, using the correlation between far-IR luminosity and dust temperature. For more details see the paper by \citet{c20} which describes the method. We apply this method to the Herschel and SMA fluxes that are available for our sources. The results of this are shown in Table \ref{table:photoz}. For the two cases where multiple SMA sources are coincident with the Herschel position (J022425 and J022448 - the second component in J021856, J021856\_2,  is too distant to make a significant contribution to the Herschel flux of the primary identification), we combine the SMA fluxes to produce an MMPz photometric redshift for the integrated source. This comes with the implicit assumption that the sources are all at a similar redshift and are physically associated. 

We also split the Herschel fluxes by assigning them to the separate SMA components based on the ratio of their SMA fluxes. This reduces the derived photometric redshifts somewhat since these divided sources are less luminous and thus are likely to have lower dust temperatures. The splitting also increases the redshift uncertainties. See Table \ref{table:photoz} for details.

\begin{table*}
\begin{tabular}{ccccccc} \hline
Source& MMPz Photoz& MMPz Range&EAZy& Zou et al. &Spec\_$z$ &Notes\\ \hline
\\
Integrated Sources\\
\\
J022425&2.73&2.1 -- 3.3\\
J022448&3.0&2.6 -- 3.5\\

\\
Split Sources\\
\\
J021856\_1&2.35&1.52 -- 3.17&0.5455&0.489$^{+0.156}_{-0.014}$&0.633&Likely lensed\\
J021856\_2&3.1&2.2 - 4.4&0.087&0.116$^{+0.317}_{-0.055}$&&Cigale fit gives $z=3.16\pm0.1$\\\\
J022425\_1&2.6&2.0 -- 3.2&2.06&1.239$^{+0.139}_{-0.05}$\\
J022425\_2&2.5&1.9 -- 3.2&2.16&2.203$^{+0.239}_{-0.277}$\\\
J022427&3.4&2.9 -- 3.8&0.4898&&0.4956&Likely lened\\
J022448\_1&2.8&2.3--3.2&2.58&2.987$^{+0.14}_{-0.897}$\\
J022448\_2&2.7&2.3 -- 3.2&2.78&2.87$^{+0.175}_{-0.311}$\\
J022448\_3&2.7&2.3 -- 3.2&2.87&2.909$^{+0.18}_{-0.289}$\\
J022610\_1&2.5&2.1 -- 3.0&&&&Cigale fit gives $z=2.52\pm$0.5\\
\\ Additional Sources\\
\\
J021856\_A1&&&1.02&1.088$^{+0.184}_{-0.112}$\\
J022427\_A1&&&2.76&2.948$^{+0.783}_{-0.093}$\\
J022448\_O1&&&2.97&2.987$^{+0.116}_{-0.086}$\\ \hline

\end{tabular}
\caption{Photometric redshifts for our targets. Results from far-IR/submm only from MMPz, and optical/NIR from EAZy from \citet{n17} and from optial/NIR with updated analysis from \citet{z22}. For MMPz the 68\% confidence interval is given. Spec\_$z$, if known, refers to the optical/NIR source and comes from \citet{n17}. The additional sources do not have SPIRE data so lack an MMPz redshift. No redshifts are given for J22610\_2 since it is only detected by the SMA. It thus lacks Herschel data to allow for an MMPz fit and also lacks optical-NIR data to allow for an EAZY fit. $^*$ This photoz is derived from CIGALE analysis of the optical to far-IR data for this source due to the paucity of optical-NIR detections.}
\label{table:photoz}
\end{table*}

\subsection{Optical/NIR Photo-z fitting}

All the sources in Table \ref{table:opt_phot}, with the exception of J022610\_1 and J022610\_2, have photometric redshifts calculated using the EAZY code and the TRACTOR photometry listed in \citet{n17}. These values are listed in \ref{table:photoz}. Furthermore, two of the optical sources identified as associated with the 500 riser Herschel sources have spectroscopic redshifts listed in \citet{n17}. These are J021856\_1, for which the spectroscopic redshift comes from the PRIMUS survey \citep{c11}, and 
the other is J022427\_1, for which the spectroscopic redshift comes from the VIPERS survey \citep{s18}. For these two sources the spectroscopic redshifts, which agree well with the EAZy optical/NIR photometric redshifts, do not agree with the MMPz derived far-IR/submm phtometric redshifts. It is also worth noting that these two sources have the largest offsets between optical/NIR positions and SMA submm positions. The reason for this disagreement is discussed below.

Of the two sources that lack an EAZY photometric redshift in the \citet{n17} catalog, J022610\_2 lacks any optical/NIR identification so a photometric redshift is impossible to obtain. J022610\_1 is only detected in three bands, Ks, and IRAC [3.6] and [4.5]. We calculate our own optical/NIR photo-z using CIGALE,  assuming reasonable upper limits in the other optical/NIR bands, and using the same models and parameter ranges as \citet{z22}. This results in a value of 3.3$\pm$0.8.

All of these sources, with the exception of J022610\_1 and \_2, also have photometric redshifts and source properties derived from EAzY using updated photometry and analyses in \citet{z22}. This paper uses data from X-ray to far-IR to derive galaxy properties for sources in the Vera Rubin Observatory Legacy Survey of Space and Time (LSST) deep drilling fields, which include the XMM field studied here. They use the same Tractor derived forced photometry approach in the optical and near-infrared as in \citet{n17} but include updated photometry that covers 13 bands from $u$ to 4.5 $\mu$m. They use spectroscopic redshifts if available, and, if not, use photometric redshifts derived by applying EAzY to this catalog. These values are listed in Table \ref{table:photoz} as `Zou et al'.  They also use CIGALE fitting in a two step process to derive the properties of these sources. For this purpose they use the optical-IR derived photometric redshifts and include photometry at longer wavelengths from the HELP project \citep{s19, s21} which uses the XID+ tool \citep{h17} to deblend far-IR fluxes from Herschel using 24$\mu$m source positions as priors since they lack the precise cross matching made possible with interferometry for the sources in this paper. 

\section{The Nature of 500 Risers}

We here assess the nature of the sources observed in this project, and specifically their multiplicity, whether physical or line of sight, and the possible role of gravitational lensing in each case. This assessment also includes a comparison of the various different photometric redshift estimations made using data in different wavelength regimes.

\subsection{Notes Individual Sources}

{\bf J021865\_1:} This source has strong and clear detections in the Herschel and SMA bands which reveal a far-IR SED that peaks around 500$\mu$m. This is suggestive of a high redshift source, confirmed by the MMPz photo-z which indicates a redshift of 2.35, albeit with large uncertainties. The optical counterpart of this source, however, is a comparatively bright object with fluxes that are inconsistent with a high redshift since it is clearly detected in the U band. Optical-NIR only photo-z estimation give values of $\sim$0.5, and there is a spectroscopic redshift of 0.633 from the PIMMS survey. This is clearly inconsistent with the redshift indicated by the far-IR/submm fluxes. There is a 1.6 arcsecond offset between the submm source seen by the SMA and the optical source (see Figure \ref{fig:sma_opt}) which is suggestive of lensing. If the optical source at $z=0.633$ is responsible for the far-IR/submm emission, this would imply an unusually low dust temperature of $\sim$10K and an unusually high dust mass. We consider this unlikely, and instead conclude that a background far-IR/submm luminous source at $z\sim 2.35$ is being lensed by the foreground, optically detected, galaxy. Lensing is common among the brightest Herschel sources but its prevalence among fainter sources such as those in the current work is unclear. Higher resolution mm/submm imaging will be needed to confirm whether lensing is in fact taking place in this object, but for the purposes of this paper we classify it as likely lened.

\begin{figure}
\includegraphics[width=8cm]{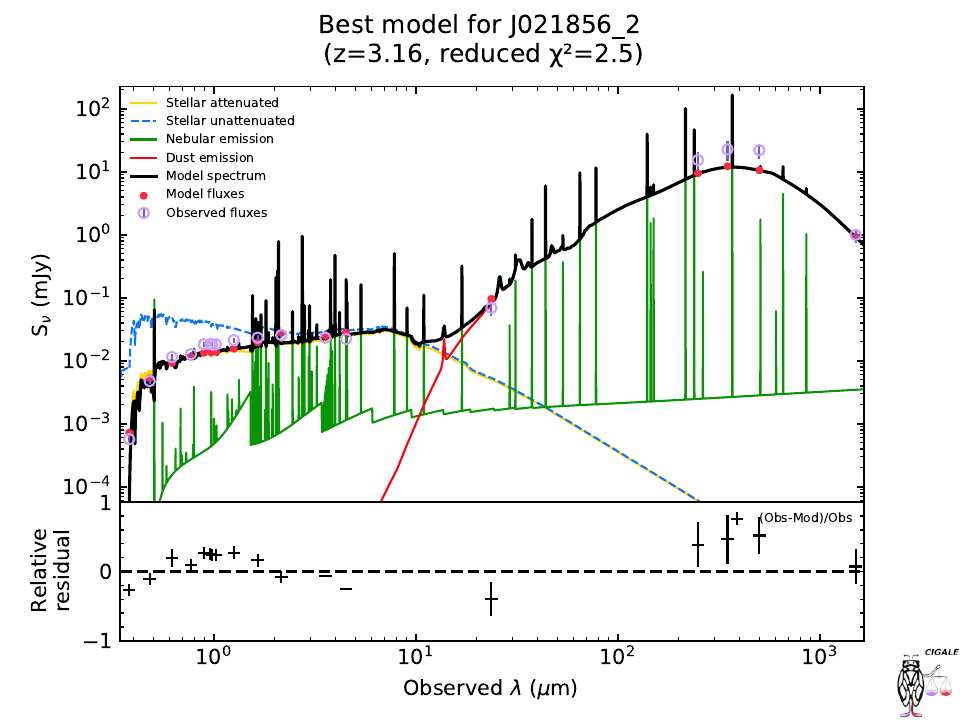}
\includegraphics[width=8cm]{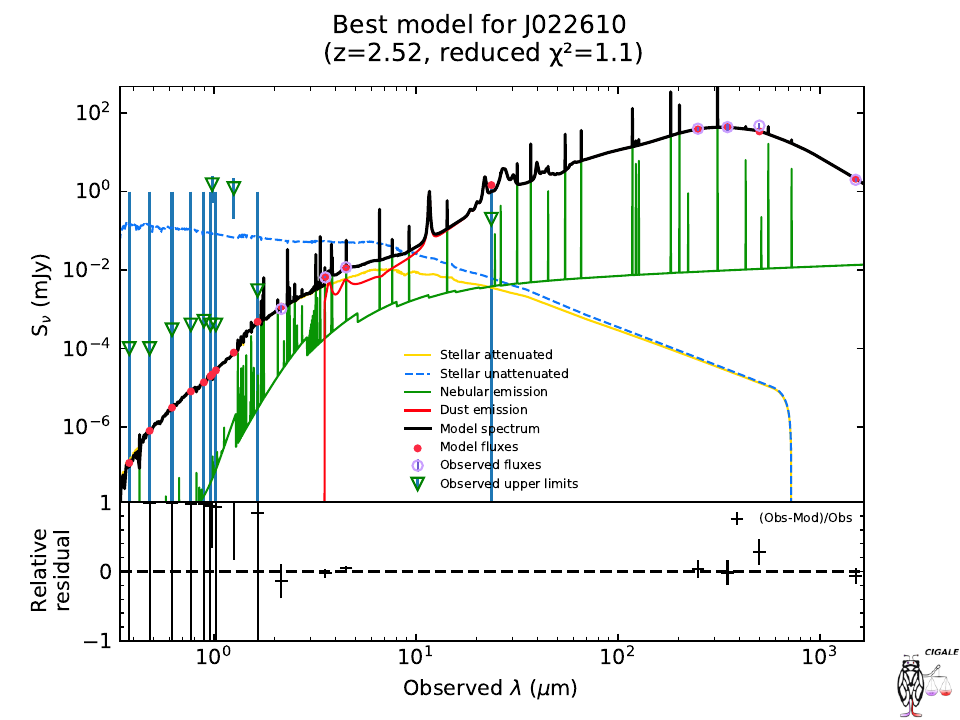}
\caption{CIGALE fits to the optical-to-mm SED for J021856\_2 and J022610. {\bf Top:} This shows that the full SED of J021856\_2 can be well fitted with a high redshift solution consistent with the MMPz photometric redshift estimate, but at much higher redshift than the optical-to-IR photo-z estimates from \citet{n17} and \citet{z22} using EAzY. {\bf Bottom:} This shows our fit to the SED of J021856\_2 which lacks optical and near-IR detections and so is absent from the 
\citet{n17} and\citet{z22} fits. As can be seen there are numerous flux upper limits in the optical to near-IR, with detections only from K' and longer wavelengths.}
\label{fig:j021856_2}
\end{figure}

{\bf J021865\_2:} This source is a serendipitously detected source offset from J021865\_2 by 23 arcseconds. It is thus unlikely to contribute significantly to the Herschel flux of J021865\_1, though it may be marginally detected by Herschel at 350$\mu$m (see Figure \ref{fig:sma_herschel_images}, flux of 19$\pm$7mJy). We use this marginal detection together with the SMA detection and flux constraints from Herschel at 250 and 500 $\mu$m to set MMPz redshift constraints which are listed in Table \ref{table:photoz}. There is substantial disagreement between the MMPz photometric redshift and that given in \citet{z22} and \citet{n17} on the basis of optical and near-IR data. If the lower redshift values were correct then this source would have an unreaosnably cold dust temperature. We thus use \verb|cigale| in photometric redshift mode to attempt a fit to the whole SED, from optical-to-mm. This produces a reasonable fit ($\chi^2$ = 2.5, compared to the Zou et al. fit which has $\chi^2$ = 4.1 for just the optical-to-near-IR) with a photometric redshift of 3.16 (see Figure \ref{fig:j021856_2}), consistent with the MMPz value and with the optical-to-IR well fitted. We suggest that this source is likely to be at high redshift and that the low redshift from \citet{z22} is an example of a catastrophic photo-z failure. The stellar mass and star formation rate derived from these fits are 6.6$\pm0.7 \times 10^{10}$M$_{\odot}$ and 900$\pm$40 M$_{\odot}$/yr respectively, which we adopt for future analysis. There remains the possibility of gravitational lensing of a background far-IR source by a foreground low redshift galaxy, but we consider this unlikely because of the small separation between SMA and optical position (0.6 arcseconds). Any ambiguity over the nature of this source can only be settled through further observations and ideally the determination of a spectroscopic redshift.

{\bf J022425\_1:} This source is the brighter of a pair of SMA sources separated by $\sim$7.2 arcseconds. They both contribute to the flux of the associated Herschel source. The photometric redshift estimates for this source are broadly consistent with a redshift of $z\sim 2$, though the \citet{z22} estimate is significantly lower.  The EAZy and MMPz estimates are consistent with this source being physically associated with J022425\_2. We thus consider it part of a potential physical pair, though this is not a secure conclusion. Spectroscopic observations will be needed to confirm this.

{\bf J022425\_2:} This source is the fainter of a pair of SMA sources separated by $\sim$7.2 arcseconds. They both contribute to the flux of the associated Herschel source. The photometric redshift estimates for this source are broadly consistent with a redshift of $z\sim 2$ which is broadly consistent with most of the redshift estimates for its companion (see above) so we conclude that this is a potential physical association. Spectroscopic observations will be needed to confirm this.

{\bf J022427\_1:} This is our brightest SMA source, detected at over 14$\sigma$. It has a distant serendipitously detected source in the same SMA field, but this lies 40 arcseconds away so will not contribute any flux in the Herschel beams. As with J021856\_1, the MMPz photometric redshift indicates a large redshift for this source. However, the optical only fit with EAZy finds a low redshift solution of 0.4898, and the \citet{z22} catalog indicates a spectroscopic redshift for the underlying optical source of 0.4956. Comparison of the optical image and SMA contours (Figure \ref{fig:sma_opt}) shows that the SMA source  is in fact offset from the optical counterpart by $\sim$2.4 arcseconds. This, together with the cold far-IR-submm SED, is highly suggestive of gravitational lensing of a background far-IR source by a foreground galaxy at the spectroscopic redshift. Higher angular resolution submm imaging will be needed to confirm the lensed nature of this source, but we classify this source as likely to be lensed.

{\bf J022448\_1, 2, 3, O1:} The source J022448\_1, and its three associated sources J022448\_2, J022448\_3 and J022448\_O1, make up a complex group of sources spread over a region of about 12 arcseconds in extent (see Figure \ref{fig:sma_opt}). All photometric redshifts for this source, as with its companions, are in broad agreement, indicating a redshift of 2.7 to 3.1. The close angular association of these sources together with the similar photometric redshifts is strongly suggestive that these four sources are physically associated, with the Herschel source clearly made up of multiple - at least three - subcomponents. The physical size of this system would be $\sim$80 kpc at $z\sim 3$ which is comparable to the size of the far-IR luminous protocluster cores found by \citet{o18} and \citet{m18} at redshifts $\sim$4. Identifying a comparable system at somewhat lower redshift potentially paves the way for studies of how protocluster cores evolve from a strongly starforming state, as seen here, to the quiescent, quenched cores found at redshifts $1.5 < z < 2.5$ (eg. \citet{w20}). We classify these sources as a physically associated multiple.

\begin{figure}
\begin{tabular}{cc}
\includegraphics[width=4cm]{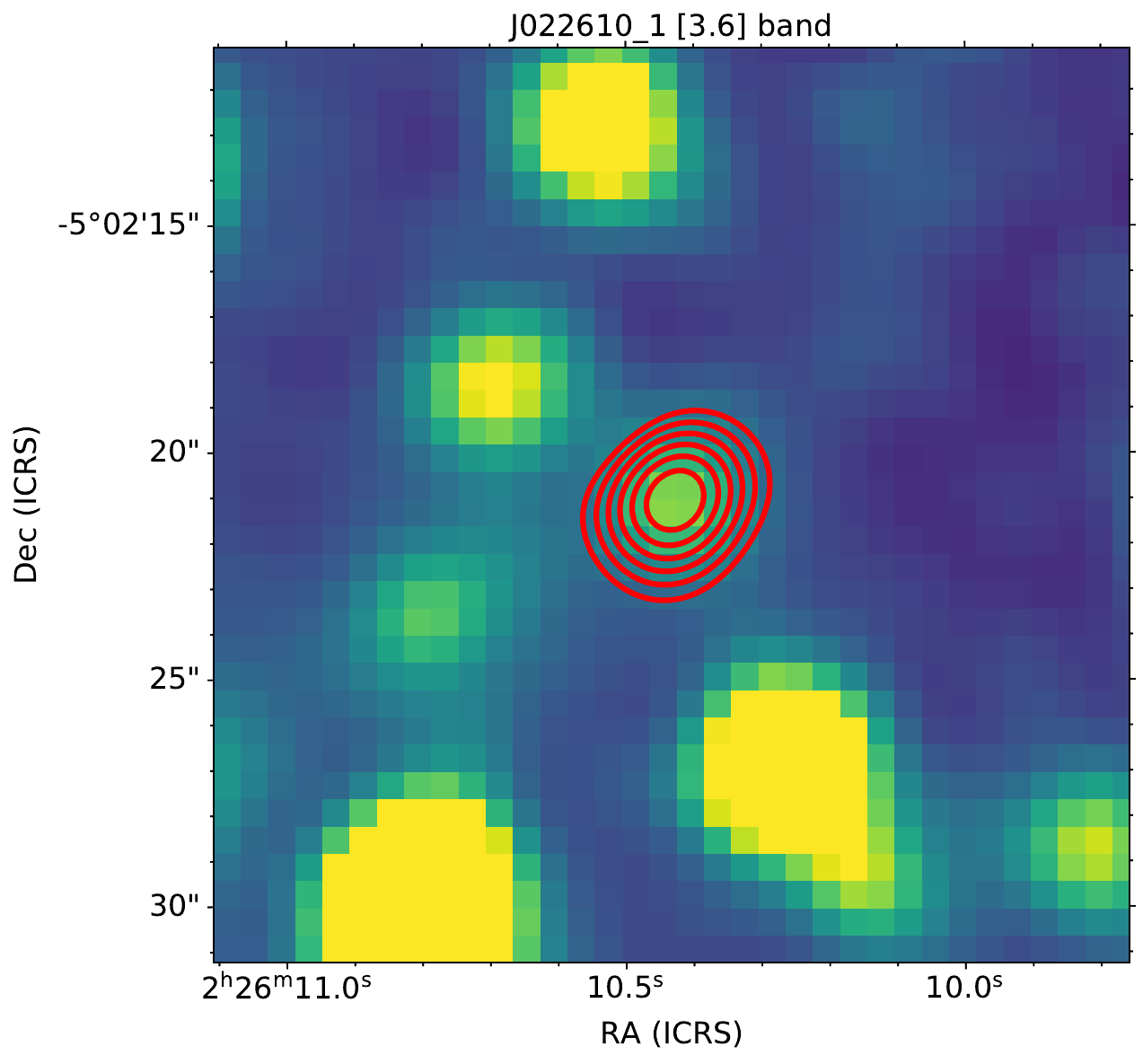} &
\includegraphics[width=4cm]{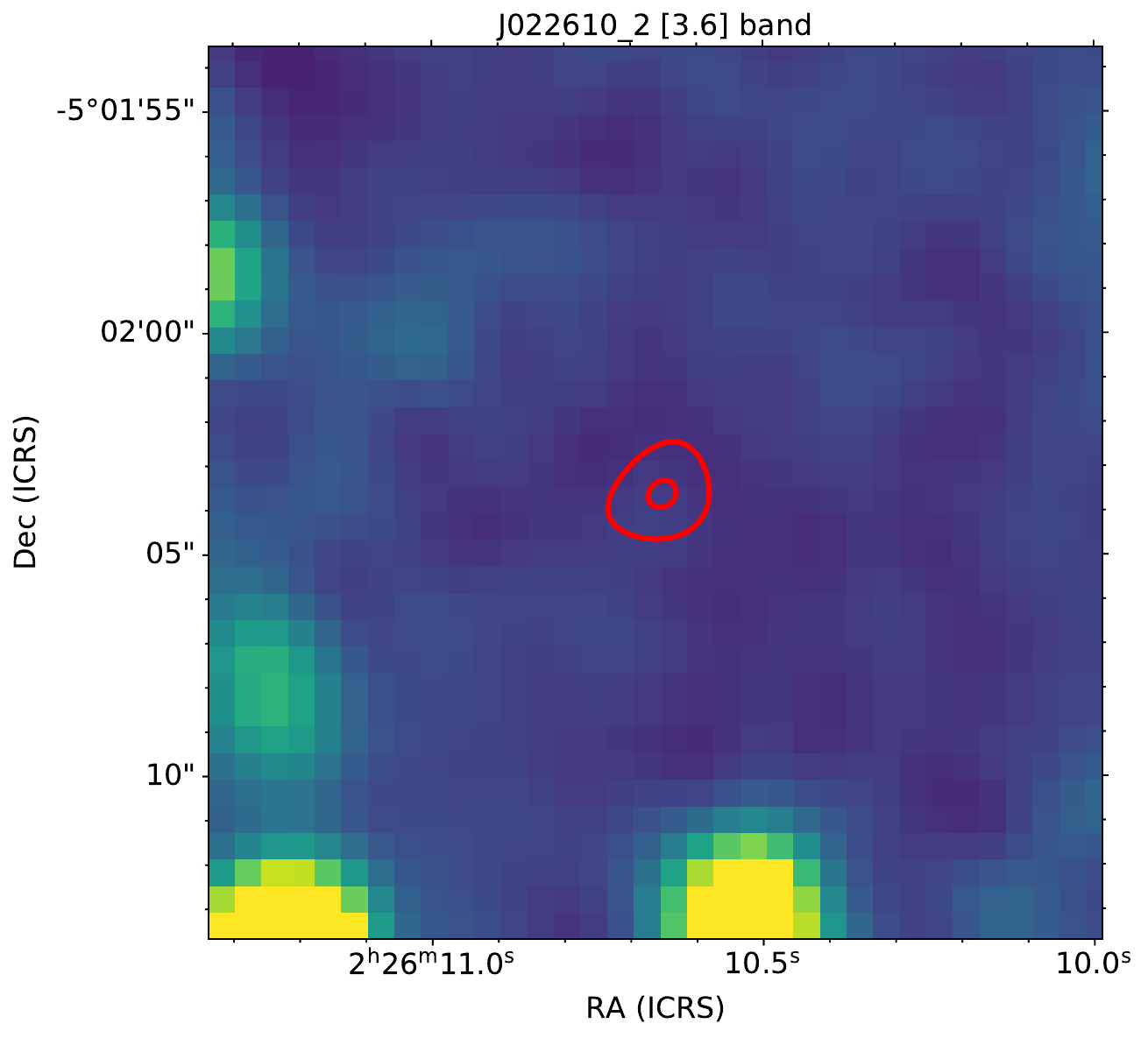} 
\\
\end{tabular}
\caption{IRAC 3.6$\mu$m images of the two 500 risers J022610\_1 and J022610\_2, not detected at shorter wavelengths, with SMA contours overlaid. SMA contours are in terms of signal to noise, starting at 3$\sigma$ and rising in increments of 1$\sigma$. As can be seen J022610\_1 is clearly detected, but J022610\_2 remains undetected at 3.6$\mu$m. They are thus both surprisingly red sources.}
\label{fig:reddest}
\end{figure}

{\bf J022610\_1:} This source is the faintest and reddest detected source in our cross identifications. It lacks detections in all bands blueward of K and has strong K-[3.6] and [3.6]-[4.5] colours. It is nevertheless strongly detected at 2mm, being our second brightest individual source with an 8.5 $\sigma$ detection. The scant optical-NIR data precludes a photometric redshift estimate from EAZy and the source does not appear in the \citet{z22}) catalog. MMPz and our own CIGALE SED fitting (see Fig. \ref{fig:j021856_2}) suggest a redshift of $z= 2.5\pm0.5$. Its derived stellar mass and star formation rate are 1.5$\pm 0.5 \times 10^{11}$M$_{\odot}$ and 1600 $\pm$ 700 M$_{\odot}$/yr respectively, which we adopt for future analysis.The source is isolated, with the nearest other SMA source lying 17.5 arcseconds away. However, that source is also optically faint, with no detections at all from U band to [4.6] (see below). 
Additional observations of this particular source are warranted by its extreme optical faintness and its brightness in the SMA observations. Its 3.6$\mu$m band image, together with SMA contours, is shown in Figure \ref{fig:reddest}

{\bf J022610\_2:} This source is a serendipitously detected source offset from J022610\_1 by 17.5 arcseconds. It is thus unlikely to contribute significantly to the Herschel flux of J022610\_1. The source lacks detections in any of our optical to NIR bands and there is no indication that the source is detected by Herschel. Its SMA flux is comparable to several other sources in this study that are detected in other bands including optical/NIR and Herschel. If its SMA-to-IRAC colours were the same as J022610\_1 it would have been detected by IRAC, so it is clearly an exceptionally red source. We can say little further about its nature given that it is only detected by the SMA, but this unusual source clearly warrants further study. The IRAC 3.6 $\mu$m image, demonstrating this non-detection, is shown in Figure \ref{fig:reddest}.

{\bf J021856\_A1:} This is a weak serendipitously detected source 20 arcseconds from J021856\_1 and 15 arcseconds from J021856\_2. We classify it as an additional source despite the low SMA S/N of 3.6$\sigma$ since it is clearly cross identified with an optical source, suggesting that it is a genuine weak detection and not a random noise source. Photometric redshifts from EAzY and \citet{z22} are consistent with a redshift of $z\sim1$. This is very different from the photometric redshifts of the background lensed source in J021856\_1 ($z\sim 2$), the foreground lens (Spec\_$z$=0.633), or the secondary source J021856\_2 ($z\sim 3$), so we conclude that there is no physical association and that its detection is genuinely serendipitous.

{\bf J022427\_A2:} This is a serendipitously detected source 40 arcseconds from J022427\_1. We classify it as an additional source despite the low SMA S/N of 3.8$\sigma$ since it is clearly cross identified with an optical source, suggesting that it is a genuine weak detection and not a random noise source. The photometric redshift estimates for this source agree on a redshift of $z \sim 2.5 - 3$. Photometric redshift estimates for J022427\_1 suggest that the lensed far-IR source is likely to lie at a higher redshift, while the foreground lens is found to be at a much lower redshift. We thus conclude that there is no physical association between these sources and that its detection is genuinely serendipitous.

\subsection{Multiplicity and lensing in 500 Risers}

We summarise our classifications of the five 500 riser Herschel sources in this complete sample in Table \ref{table:classifications}. We classify sources as single or multiple in terms of the contribution of sources to the Herschel flux. Association indicates whether another source nearby is likely to be physically associated, even if it is too far away to contribute to the detected Herschel flux.

\begin{table}
\begin{tabular}{ll} \hline
Source&Classification\\ \hline
J021856 & Single source, Gravitationally Lensed\\
J022425& Multiple source, possible physical pair\\
J022427& Single source, Gravitationally lensed\\
J022448& Multiple source, physically associated\\
J022610& Single source, possible association with a distant companion\\ \hline
\end{tabular}
\caption{The nature of the Herschel 500 risers as revealed by our SMA observations. Single source means that one object is likely responsible for the Herschel flux of the 500 riser, multiple means that more than one SMA source contributes.}
\label{table:classifications}
\end{table}

Of the five sources that make up this small but complete sample of 500 risers, we find that two are likely to be gravitationally lensed, that two are multiple, and that one is a single, likely unlensed source. Of the two multiple sources, one is very likely to be a physical association, while the other is a potential physical association.

\section{Discussion}

\subsection{The Nature of 500 risers: Blends and lensing}

Theoretical expectations for the reddest {\em Herschel} sources have suggested that they should be almost entirely 
blends of multiple sources \citep{s16, scu18, c15}. This is clearly not the case here, with three of our sources - thus 60\% of this complete sample - being single sources. \citet{b17} instead suggest that S$_{500} < 60$ mJy sources will largely be multiple with the brightest source contributing only 60\% of the total flux, while brighter sources will in general be strongly lensed objects. All of our sources lie below the 60mJy level so, according to the \citet{b17} simulations, they should be multiple, and lensing should not play a significant role. We find something rather different. Of the three isolated sources in the sample, two are likely to be lensing systems, with the optical identification - the putative foreground lens - having spectroscopic redshifts highly discordant with the estimated redshift of the far-IR source. This suggests that 40\% of this complete sample of 500 risers are in fact lensed objects. This is somewhat discordant with predictions given that the redshifts of these putative lenses, and indeed the rest of this 500 riser sample, are around $z=2.5 - 3.5$. Predictions, based on the models of \citet{ne17} and discussed in the context of 500 risers observed by ALMA \citep{o17}, suggest that the lensing fraction of S$_{500} = 40 - 60$ mJy sources rises with increasing redshift, and thus redder colour. The lensing fraction for our sources should be $\sim$20\% rather than the 40\% seen here (see Figure 2 in \citet{o17}). The limitations of small number statistics apply, with a definitive test of whether the models are failing to match the data requiring a larger complete sample than is available here.

The possibility exists that some of our sources are in fact multiple but that the majority of the components are too faint for us to detect in our SMA imaging. This could indeed be the case for one of our single isolated sources, the putative lensed source J021856. This is detected by the SMA at just over 5$\sigma$ significance, so companion sources only a small factor fainter would not have been detected. Our other single sources, however, are detected at much higher significance, with 
J022427 detected at 14.2 $\sigma$ and J022610 detected at 8.5 $\sigma$ putative companions a factor of 2 to 3 times fainter would thus still have been detected in these cases. These sources are thus only consistent with the \citet{b17} prediction that the brightest contributor to a blended 500 riser contributes only 60\% of the total flux if there are multiple faint companion sources. Of our multiple sources, J022425 and J022448, J022425 is roughly consistent with the prediction, with the brighter of the two sources contributing 57\% of the total flux, assuming there are no undetected fainter companions. J022448, in contrast, consists of three sources of roughly equal brightness, with the brightest contributing slightly over a third of the flux. These observations suggest that the lensing rate and multiplicity of 500 risers is rather more complex than currently predicted. Deeper interferometric observations could address this issue, but, as shown by \citet{c23}, it can also be investigated by comparing the resolved fluxes for these sources from the SMA with unresolved fluxes measured by a submm camera such as SCUBA2. 

Comparing our results to other observations of 500 risers, from \citet{o17}, \citet{g20} and \citet{c23} we perhaps find better agreement, with the 500 riser populations found in these various large but statistically inhomogeneous studies. \citet{o17} find that 17/44 500 risers in their sample are multiple, matching the 40\% found here. \citet{g20} find that $\sim$ 35\% of sources brighter than 60mJy at 500$\mu$m are blends with a larger proportion than predicted at fainter fluxes, though not all of their targets were detected. Followup observations of four of these undetected sources by \citet{c23} found that three of them were multiple, raising the multiplicity fraction somewhat.  The one source in \citet{c23} that was not multiple is one of the isolated sources in the current paper, J021856. The multiplicity rate of the small complete sample described here is thus consistent with the multiplicity rate found in other observations, though the sources discussed here are at the faint end of the 500$\mu$m flux distributions discussed in these past papers. The lensing rate in other studies is less well defined. The lensing candidates described here are identified on the basis of a disagreement between the far-IR and optical-NIR photometric redshifts, while the possibility of lensing in the other studies is based on submm morphologies. This is especially the case for the \citet{o17} sources which were observed by ALMA at a resolution of 0.14 arcseconds, allowing arc and ring-like features to be used to indicate lensing as well as proximity to near-IR counterparts in the absence of explicit lensing features. They find 18/44 500 risers to be lensed, a rate of 41\% consistent with the 40\% rate found for the complete sample discussed here, albeit for a somewhat fainter and less red sample of objects. A lower fraction of lensing was found in \citet{g20}, but their angular resolution was in general worse then either \citet{o17} or the current work, so this number should be regarded as a lower limit. Lower resolution 1.1mm observations of 500 risers with the LMT \citep{m21} found that 9\% were multiple on scales larger than 9 arcseconds. The multiple sources observed here separations smaller than this so the low multiple fraction seen in this data is consistent with our observations. Observations of 500 risers with IRAC at 3.6 and 4.5$\mu$m by \citet{m19} found that 65\% were unlensed, consistent with our lensing rate of 40\%, but found a somewhat lower multiplicity rate of 27\% compared to our rate of 40\%, though these numbers are consistent given the small number of sources in our sample.

In summary, observations of 500 risers, including this work, find that about 40\% of them are blends, about 40\% are lensed, and the remaining 20\% are isolated single sources.

\subsection{Physical Associations}

Given that a significant fraction of 500 risers turn out to be blends of two or more sources within the large {\em Herschel} beams, the question then arises as to whether these blends are the result of chance line of sight alignment or whether they are genuine physical associations. For the two blended sources identified in the sample discussed here, we assess the possibility of physical association on the basis of angular separation and similarity of the photometric redshifts of the contributing sources. Of our two potential multiple sources, J022425 is a plausible physical association. The angular separation is 7.2 arcseconds and most of the redshift estimates are consistent with each other. The \citet{z22} CIGALE based redshift estimates, however, are significantly different so there remains the possibility that this source is just a chance alignment. A solid redshift determination is needed to be sure. Our other multiple source, J022448, is rather different and is very likely a physical association with small angular separations and largely consistent redshift estimates within the errors. There is also a fourth component, selected as an optical companion, which lies at the same redshift. We thus believe that J022448 is very likely a genuine physical association and may be lower redshift equivalent to the Deep Red Core discovered by \citet{o18} at $z=4.002$. We discuss this system in more detail below.

\subsection{Star Formation in 500 risers}

We can use the SED fits from \citet{z22} and those described here to examine the properties of these galaxies. In particular we can use the stellar mass and star formation rate estimates from these SED fits to determine where our 500 risers and associated objects sit on the stellar-mass-SFR correlation, sometimes referred to as the galaxy `main sequence' (eg. \citet{sp14}). The positions of our sources in this diagram together with the galaxy `main sequence' at a redshift of 3 are shown in Figure 6. As can be seen most of our sources do indeed lie on the main sequence. Two sources lie below it, however, having substantially less star formation than their stellar mass would suggest, implying that they are quenched objects. One of these is J022427\_1, a putative lensing galaxy. The quenched status of this source is entirely consistent with its being a foreground quenched elliptical lensing a background far-IR luminous DSFG, as seen in other examples of lensed Herschel galaxies (eg. \citet{n10}). The other quenched galaxy is the companion optical source to the likely physical group of SMA sources associated with J022448, and suggests that this object is a massive quenched galaxy at a redshift $\sim$3. We discuss this further in the next section.

Most of the rest of the sources lie on or somewhat above the main sequence, suggesting they are largely normal star forming galaxies, with star formation rates of $\sim$100 to $\sim$1000 M$_{\odot}$/yr, consistent with other observations of 500 risers (eg. \citet{m19}). A few sources lie somewhat above the main sequence suggesting the presence of starbursts. These include the serendipitously detected source J021865\_2, the source lacking optical counterparts, J022610\_1, which has our highest star formation rate of 1600 M$_{\odot}$/yr, and one of the sources within the J022448 association. We thus conclude that 500 risers are generally normally star forming galaxies at their appropriate redshift but that a significant fraction are, or contain, active starbursting galaxies.

\begin{figure}
\includegraphics[width=9cm]{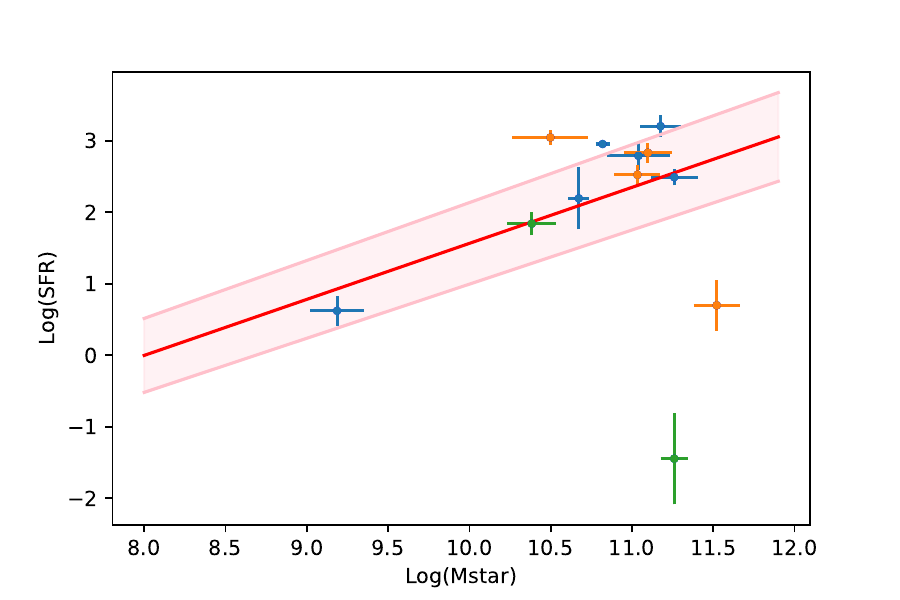}
\label{fig:ms}
\caption{Location of our sources on the star-formation rate - stellar mass correlation. The red line indicates the position of this correlation for a redshift of 3 \citep{sp14}, while the data points show the position of our sources on this diagram. Different coloured points represent different subsamples. Green points represent the sources identified as lens candidates. Orange points indicate the sources associated with the putative galaxy cluster core associated with source J022448. The rest of the sources are shown as blue points. Stellar masses and star formation rates come from the SED fits in \citet{z22} or from our own fits for J021856\_2 and J022610.}
\end{figure}

\subsection{The J022448 Cluster Candidate}

The most unusual source in our small complete sample is the multiple sources associated with J022448. This consists of three SMA detected submm sources and a nearby optical source within a region $\sim$7 arcseconds across, all lying at photometric redshifts $z\sim 2.5 - 3$. The sources are thus likely to all be physically associated, making this a candidate starbursting cluster core, similar to those identified by \citet{o18} at $z=4.002$ and \citet{m18} at $z=4.3$, though at a somewhat lower redshift. However, unlike the galaxies found in these higher redshift systems \citep{lo20,ro21}, one of these sources sits well below the star formation `main sequence', appearing to be roughly 2 order of magnitude deficient in star formation for its stellar mass. This compares to the most quenched galaxy in the $z=4.3$ cluster core lying about 0.75 dex below the main sequence. This particular source, J022448\_O1, has a CIGALE derived star formation rate, derived from the optical/near-IR data and assuming zero flux in the far-IR and submm, of 5$\pm$6 M$_{\odot}$/yr and a stellar mass of 3.3$\pm$ 1.2 M$_{\odot}$. The high stellar mass and low star formation rate make this source a potential candidate brightest cluster galaxy (BCG) that has already quenched but which is still accreting star-forming companion galaxies. As such this system could be a valuable probe into somewhat later stages in the evolution of a BCG and cluster core than are seen in the Otero and Miller objects at higher redshift. However, further observations, and especially spectroscopic confirmation that these are genuinely physically associated, is needed before these conclusions are secure.

\section{Conclusions}

We have conducted SMA observations of a small but complete sample of 500 risers in the XMMLSS region to allow us to examine the optical-to-far-infrared properties of these sources so as to better understand the nature of these sources. Out of our complete sample of five sources we find that two are likely to be gravitationally lensed, two are likely to be in multiple systems blended together by the {\em Herschel} beam, and one is an isolated single source. The rates of lensing and multiplicity that we find are compatible with those found in other, heterogeneously selected 500 riser samples. Photometric redshift estimates for the far-IR sources like in the range $z \sim 2.5 - 3$, though the putative lenses for the two lensed systems lie at considerably lower redshifts, $z \sim 0.5$. These results are also consistent with earlier work. We thus conclude that 500 risers are a more diverse population than some of the theoretical models have suggested, with a mixture of lensed, multiples and isolated sources.

We examine the properties of our sources using CIGALE fits to their observed optical-to-far-IR SEDs and find that most of the sources lie on the star-forming main sequence, though with a few sources lying above it can be classified as starbursts.

One of our multiple sources, J022448, appears to be a physical association of three far-IR sources and an addition massive, far-IR faint, optically identified galaxy all at $z\sim3$. We suggest that this group of sources is the core of a forming galaxy cluster with the massive optical source, which lies well below the star-forming main sequence, being a BCG in the process of accreting actively star forming neighbours. Further observations of this system will provide new insights into the assembly of BCGs and the triggering and quenching of star formation during this process.
\\~\\

{\bf Acknowledgements}
The authors wish to recognise and acknowledge the very significant cultural role and reverence that the summit of Mauna Kea has always had within the indigenous Hawaiian community.  We are most fortunate to have the opportunity to conduct observations from this mountain.
The Submillimeter Array is a joint project between the Smithsonian Astrophysical Observatory and the Academia Sinica Institute of Astronomy and Astrophysics and is funded by the Smithsonian Institution and the Academia Sinica.
This research has made use of data from the HerMES project (http://hermes.sussex.ac.uk/). HerMES is a Herschel Key Programme utilising Guaranteed Time from the SPIRE instrument team, ESAC scientists and a mission scientist.
The HerMES data was accessed through the Herschel Database in Marseille (HeDaM - http://hedam.lam.fr) operated by CeSAM and hosted by the Laboratoire d'Astrophysique de Marseille.
This research made use of Astropy, a community-developed core Python package for Astronomy \citep{astropy}.
and of the Starlink Table/VOTable Processing Software, TOPCAT \citep{Taylor2005}.
This research also made use of NASA's Astrophysics Data System Bibliographic Services.
DLC, JC and JG acknowledge support from STFC, in part through grant numbers ST/N000838/1 and ST/K001051/1. DR acknowledges support from the National Science Foundation under grant number AST-1614213 and from the Alexander von Humboldt Foundation and the Federal Ministry for Education and Research through a Humboldt Research Fellowship for Experienced Researchers. IPF acknowledges support from the Spanish Ministerio de Ciencia, Innovacion
y Universidades (MICINN) under grant numbers ESP2015-65597-C4-4-R and ESP2017-86852-C4-2-R. Thanks to Caitlin Casey for providing the MMPZ code. Many thanks to Tai-An Cheng and Mattia Negrello for useful comments.
\\~\\
{\bf Data Availability}
The data for this paper is available through the SMA observatory archives at https://lweb.cfa.harvard.edu/sma-archive/ and from relevant references.
\\~\\


\begin{thebibliography}{99}
\bibitem[\protect\citeauthoryear{Aihara et al.}{2018}]{a18} Aihara H., Arimoto N., Armstrong R., Arnouts S., Bahcall N.~A., Bickerton S., Bosch J., et al., 2018, PASJ, 70, S4. doi:10.1093/pasj/psx066
\bibitem[\protect\citeauthoryear{Asboth et al.}{2016}]{a16} Asboth V., Conley A., Sayers J., B{\'e}thermin M., Chapman S.~C., Clements D.~L., Cooray A., et al., 2016, MNRAS, 462, 1989. doi:10.1093/mnras/stw1769
\bibitem[\protect\citeauthoryear{Astropy Collaboration et al.}{2022}]{astropy} Astropy Collaboration, Price-Whelan A.~M., Lim P.~L., Earl N., Starkman N., Bradley L., Shupe D.~L., et al., 2022, ApJ, 935, 167. doi:10.3847/1538-4357/ac7c74
\bibitem[\protect\citeauthoryear{Baugh et al.}{2005}]{b05} Baugh C.~M., Lacey C.~G., Frenk C.~S., Granato G.~L., Silva L., Bressan A., Benson A.~J., et al., 2005, MNRAS, 356, 1191. doi:10.1111/j.1365-2966.2004.08553.x
\bibitem[\protect\citeauthoryear{B{\'e}thermin et al.}{2017}]{b17} B{\'e}thermin M., Wu H.-Y., Lagache G., Davidzon I., Ponthieu N., Cousin M., Wang L., et al., 2017, A\&A, 607, A89. doi:10.1051/0004-6361/201730866
\bibitem[\protect\citeauthoryear{Cairns et al.}{2023}]{c23} Cairns J., Clements D.~L., Greenslade J., Petitpas G., Cheng T., Ding Y., Parmar A., et al., 2023, MNRAS, 519, 709. doi:10.1093/mnras/stac3486
\bibitem[\protect\citeauthoryear{Casey}{2020}]{c20} Casey C.~M., 2020, ApJ, 900, 68. doi:10.3847/1538-4357/aba528
\bibitem[\protect\citeauthoryear{Chapman et al.}{2005}]{c05} Chapman S.~C., Blain A.~W., Smail I., Ivison R.~J., 2005, ApJ, 622, 772. doi:10.1086/428082
\bibitem[\protect\citeauthoryear{Coil et al.}{2011}]{c11} Coil A.~L., Blanton M.~R., Burles S.~M., Cool R.~J., Eisenstein D.~J., Moustakas J., Wong K.~C., et al., 2011, ApJ, 741, 8. doi:10.1088/0004-637X/741/1/8
\bibitem[\protect\citeauthoryear{Cowley et al.}{2015}]{c15} Cowley W.~I., Lacey C.~G., Baugh C.~M., Cole S., 2015, MNRAS, 446, 1784. doi:10.1093/mnras/stu2179
\bibitem[\protect\citeauthoryear{Cowley et al.}{2017}]{c17} Cowley W.~I., Lacey C.~G., Baugh C.~M., Cole S., Wilkinson A., 2017, MNRAS, 469, 3396. doi:10.1093/mnras/stx928
\bibitem[\protect\citeauthoryear{Donevski et al.}{2018}]{d18} Donevski D., Buat V., Boone F., Pappalardo C., Bethermin M., Schreiber C., Mazyed F., et al., 2018, A\&A, 614, A33. doi:10.1051/0004-6361/201731888
\bibitem[\protect\citeauthoryear{Dowell et al.}{2014}]{d14} Dowell C.~D., Conley A., Glenn J., Arumugam V., Asboth V., Aussel H., Bertoldi F., et al., 2014, ApJ, 780, 75. doi:10.1088/0004-637X/780/1/75
\bibitem[\protect\citeauthoryear{Eales et al.}{2010}]{e10} Eales S., Dunne L., Clements D., Cooray A., De Zotti G., Dye S., Ivison R., et al., 2010, PASP, 122, 499. doi:10.1086/653086
\bibitem[\protect\citeauthoryear{Greenslade et al.}{2019}]{g19} Greenslade J., Aguilar E., Clements D.~L., Dannerbauer H., Cheng T., Petitpas G., Yang C., et al., 2019, MNRAS, 490, 5317. doi:10.1093/mnras/stz2850
\bibitem[\protect\citeauthoryear{Greenslade et al.}{2020}]{g20} Greenslade J., Clements D.~L., Petitpas G., Asboth V., Conley A., P{\'e}rez-Fournon I., Riechers D., 2020, MNRAS, 496, 2315. doi:10.1093/mnras/staa1637
\bibitem[\protect\citeauthoryear{Griffin et al.}{2010}]{g10} Griffin M.~J., Abergel A., Abreu A., Ade P.~A.~R., Andr{\'e} P., Augueres J.-L., Babbedge T., et al., 2010, A\&A, 518, L3. doi:10.1051/0004-6361/201014519
\bibitem[\protect\citeauthoryear{Gruppioni et al.}{2011}]{gr11} Gruppioni C., Pozzi F., Zamorani G., Vignali C., 2011, MNRAS, 416, 70. doi:10.1111/j.1365-2966.2011.19006.x
\bibitem[\protect\citeauthoryear{Hayward et al.}{2011}]{h11} Hayward C.~C., Kere{\v{s}} D., Jonsson P., Narayanan D., Cox T.~J., Hernquist L., 2011, ApJ, 743, 159. doi:10.1088/0004-637X/743/2/159
\bibitem[\protect\citeauthoryear{Henriques et al.}{2015}]{h15} Henriques B.~M.~B., White S.~D.~M., Thomas P.~A., Angulo R., Guo Q., Lemson G., Springel V., et al., 2015, MNRAS, 451, 2663. doi:10.1093/mnras/stv705
\bibitem[\protect\citeauthoryear{Heymans et al.}{2012}]{h12} Heymans C., Van Waerbeke L., Miller L., Erben T., Hildebrandt H., Hoekstra H., Kitching T.~D., et al., 2012, MNRAS, 427, 146. doi:10.1111/j.1365-2966.2012.21952.x
\bibitem[\protect\citeauthoryear{Hurley et al.}{2017}]{h17} Hurley P.~D., Oliver S., Betancourt M., Clarke C., Cowley W.~I., Duivenvoorden S., Farrah D., et al., 2017, MNRAS, 464, 885. doi:10.1093/mnras/stw2375
\bibitem[\protect\citeauthoryear{Jarvis et al.}{2013}]{j13} Jarvis M.~J., Bonfield D.~G., Bruce V.~A., Geach J.~E., McAlpine K., McLure R.~J., Gonz{\'a}lez-Solares E., et al., 2013, MNRAS, 428, 1281. doi:10.1093/mnras/sts118
\bibitem[\protect\citeauthoryear{Lacey et al.}{2010}]{l10} Lacey C.~G., Baugh C.~M., Frenk C.~S., Benson A.~J., Orsi A., Silva L., Granato G.~L., et al., 2010, MNRAS, 405, 2. doi:10.1111/j.1365-2966.2010.16463.x
\bibitem[\protect\citeauthoryear{Lacey et al.}{2016}]{l16} Lacey C.~G., Baugh C.~M., Frenk C.~S., Benson A.~J., Bower R.~G., Cole S., Gonzalez-Perez V., et al., 2016, MNRAS, 462, 3854. doi:10.1093/mnras/stw1888
\bibitem[\protect\citeauthoryear{Lang, Hogg, \& Schlegel}{2016}]{l16a} Lang D., Hogg D.~W., Schlegel D.~J., 2016, AJ, 151, 36. doi:10.3847/0004-6256/151/2/36
\bibitem[\protect\citeauthoryear{Lang, Hogg, \& Mykytyn}{2016}]{l16b} Lang D., Hogg D.~W., Mykytyn D., 2016, ascl.soft. ascl:1604.008
\bibitem[\protect\citeauthoryear{Long et al.}{2020}]{lo20} Long A.~S., Cooray A., Ma J., Casey C.~M., Wardlow J.~L., Nayyeri H., Ivison R.~J., et al., 2020, ApJ, 898, 133. doi:10.3847/1538-4357/ab9d1f
\bibitem[\protect\citeauthoryear{Lonsdale et al.}{2003}]{l03} Lonsdale C.~J., Smith H.~E., Rowan-Robinson M., Surace J., Shupe D., Xu C., Oliver S., et al., 2003, PASP, 115, 897. doi:10.1086/376850
\bibitem[\protect\citeauthoryear{Ma et al.}{2019}]{m19} Ma J., Cooray A., Nayyeri H., Brown A., Ghotbi N., Ivison R., Oteo I., et al., 2019, ApJS, 244, 30. doi:10.3847/1538-4365/ab4194
\bibitem[\protect\citeauthoryear{Martin et al.}{2005}]{m05} Martin D.~C., Fanson J., Schiminovich D., Morrissey P., Friedman P.~G., Barlow T.~A., Conrow T., et al., 2005, ApJL, 619, L1. doi:10.1086/426387
\bibitem[\protect\citeauthoryear{Mauduit et al.}{2012}]{m12} Mauduit J.-C., Lacy M., Farrah D., Surace J.~A., Jarvis M., Oliver S., Maraston C., et al., 2012, PASP, 124, 714. doi:10.1086/666945
\bibitem[\protect\citeauthoryear{Miller et al.}{2018}]{m18} Miller T.~B., Chapman S.~C., Aravena M., Ashby M.~L.~N., Hayward C.~C., Vieira J.~D., Wei{\ss} A., et al., 2018, Natur, 556, 469. doi:10.1038/s41586-018-0025-2
\bibitem[\protect\citeauthoryear{Monta{\~n}a et al.}{2021}]{m21} Monta{\~n}a A., Zavala J.~A., Aretxaga I., Hughes D.~H., Ivison R.~J., Pope A., S{\'a}nchez-Arg{\"u}elles D., et al., 2021, MNRAS, 505, 5260. doi:10.1093/mnras/stab1649
\bibitem[\protect\citeauthoryear{Negrello et al.}{2010}]{n10} Negrello M., Hopwood R., De Zotti G., Cooray A., Verma A., Bock J., Frayer D.~T., et al., 2010, Sci, 330, 800. doi:10.1126/science.1193420
\bibitem[\protect\citeauthoryear{Negrello et al.}{2017}]{ne17} Negrello M., Gonzalez-Nuevo J., De Zotti G., Bonato M., Cai Z.-Y., Clements D., Danese L., et al., 2017, MNRAS, 470, 2253. doi:10.1093/mnras/stx1367
\bibitem[\protect\citeauthoryear{Nyland et al.}{2017}]{n17} Nyland K., Lacy M., Sajina A., Pforr J., Farrah D., Wilson G., Surace J., et al., 2017, ApJS, 230, 9. doi:10.3847/1538-4365/aa6fed
\bibitem[\protect\citeauthoryear{Oliver et al.}{2012}]{o12} Oliver S.~J., Bock J., Altieri B., Amblard A., Arumugam V., Aussel H., Babbedge T., et al., 2012, MNRAS, 424, 1614. doi:10.1111/j.1365-2966.2012.20912.x
\bibitem[\protect\citeauthoryear{Oteo et al.}{2017}]{o17} Oteo I., Ivison R.~J., Negrello M., Smail I., P{\'e}rez-Fournon I., Bremer M., De Zotti G., et al., 2017, arXiv, arXiv:1709.04191. doi:10.48550/arXiv.1709.04191
\bibitem[\protect\citeauthoryear{Oteo et al.}{2018}]{o18} Oteo I., Ivison R.~J., Dunne L., Manilla-Robles A., Maddox S., Lewis A.~J.~R., de Zotti G., et al., 2018, ApJ, 856, 72. doi:10.3847/1538-4357/aaa1f1
\bibitem[\protect\citeauthoryear{Riechers et al.}{2013}]{r13} Riechers D.~A., Bradford C.~M., Clements D.~L., Dowell C.~D., P{\'e}rez-Fournon I., Ivison R.~J., Bridge C., et al., 2013, Natur, 496, 329. doi:10.1038/nature12050
\bibitem[\protect\citeauthoryear{Riechers et al.}{2021}]{r21} Riechers D.~A., Nayyeri H., Burgarella D., Emonts B.~H.~C., Clements D.~L., Cooray A., Ivison R.~J., et al., 2021, ApJ, 907, 62. doi:10.3847/1538-4357/abcf2e
\bibitem[\protect\citeauthoryear{Rotermund et al.}{2021}]{ro21} Rotermund K.~M., Chapman S.~C., Phadke K.~A., Hill R., Pass E., Aravena M., Ashby M.~L.~N., et al., 2021, MNRAS, 502, 1797. doi:10.1093/mnras/stab103
\bibitem[\protect\citeauthoryear{Rowan-Robinson et al.}{2016}]{rr16} Rowan-Robinson M., Oliver S., Wang L., Farrah D., Clements D.~L., Gruppioni C., Marchetti L., et al., 2016, MNRAS, 461, 1100. doi:10.1093/mnras/stw1169
\bibitem[\protect\citeauthoryear{Scodeggio et al.}{2018}]{s18} Scodeggio M., Guzzo L., Garilli B., Granett B.~R., Bolzonella M., de la Torre S., Abbas U., et al., 2018, A\&A, 609, A84. doi:10.1051/0004-6361/201630114
\bibitem[\protect\citeauthoryear{Scudder et al.}{2016}]{s16} Scudder J.~M., Oliver S., Hurley P.~D., Griffin M., Sargent M.~T., Scott D., Wang L., et al., 2016, MNRAS, 460, 1119. doi:10.1093/mnras/stw1044
\bibitem[\protect\citeauthoryear{Scudder et al.}{2018}]{scu18} Scudder J.~M., Oliver S., Hurley P.~D., Wardlow J.~L., Wang L., Farrah D., 2018, MNRAS, 480, 4124. doi:10.1093/mnras/sty2009
\bibitem[\protect\citeauthoryear{Shirley et al.}{2019}]{s19} Shirley R., Roehlly Y., Hurley P.~D., Buat V., Campos Varillas M. del C., Duivenvoorden S., Duncan K.~J., et al., 2019, MNRAS, 490, 634. doi:10.1093/mnras/stz2509
\bibitem[\protect\citeauthoryear{Shirley et al.}{2021}]{s21} Shirley R., Duncan K., Campos Varillas M.~C., Hurley P.~D., Ma{\l}ek K., Roehlly Y., Smith M.~W.~L., et al., 2021, MNRAS, 507, 129. doi:10.1093/mnras/stab1526
\bibitem[\protect\citeauthoryear{Shupe et al.}{2008}]{s08} Shupe D.~L., Rowan-Robinson M., Lonsdale C.~J., Masci F., Evans T., Fang F., Oliver S., et al., 2008, AJ, 135, 1050. doi:10.1088/0004-6256/135/3/1050
\bibitem[\protect\citeauthoryear{Speagle et al.}{2014}]{sp14} Speagle J.~S., Steinhardt C.~L., Capak P.~L., Silverman J.~D., 2014, ApJS, 214, 15. doi:10.1088/0067-0049/214/2/15
\bibitem[\protect\citeauthoryear{Taylor}{2005}]{Taylor2005} Taylor M.~B., 2005, ASPC, 347, 29
\bibitem[\protect\citeauthoryear{Willis et al.}{2020}]{w20} Willis J.~P., Canning R.~E.~A., Noordeh E.~S., Allen S.~W., King A.~L., Mantz A., Morris R.~G., et al., 2020, Natur, 577, 39. doi:10.1038/s41586-019-1829-4
\bibitem[\protect\citeauthoryear{Yan, Ling, \& Ma}{2022}]{y22} Yan H., Ling C., Ma Z., 2022, MNRAS, 516, 5471. doi:10.1093/mnras/stac2502
\bibitem[\protect\citeauthoryear{Zou et al.}{2022}]{z22} Zou F., Brandt W.~N., Chen C.-T., Leja J., Ni Q., Yan W., Yang G., et al., 2022, ApJS, 262, 15. doi:10.3847/1538-4365/ac7bdf

\end{thebibliography}
\end{document}